\documentclass[12pt,draftclsnofoot,onecolumn]{IEEEtran}
\makeatletter
\def\ps@headings{%
\def\@oddhead{\mbox{}\scriptsize\rightmark \hfil \thepage}%
\def\@evenhead{\scriptsize\thepage \hfil \leftmark\mbox{}}%
\def\@oddfoot{}%
\def\@evenfoot{}}
\makeatother
\usepackage{multirow}
\usepackage{amssymb}
\usepackage{amsmath}
\usepackage{amsfonts}
\usepackage{graphicx}
\usepackage{subfigure}
\usepackage{cite}
\usepackage{enumerate}
\usepackage{url}
\usepackage{bm}
\usepackage{clrscode}
\usepackage{subfigure}
\usepackage{amsthm}

\DeclareMathOperator*{\argmax}{argmax}

\begin{document}

\title{\large Sub-channel and Power Allocation for Non-orthogonal Multiple Access Relay Networks with Amplify-and-Forward Protocol}
%\author{
%\IEEEauthorblockN{
%\normalsize{Shuhang Zhang}\IEEEauthorrefmark{1},
%\normalsize{Boya Di}\IEEEauthorrefmark{1},
%\normalsize{Lingyang Song}\IEEEauthorrefmark{1},
%\normalsize{and Yonghui Li}\IEEEauthorrefmark{2},\\}
%\IEEEauthorblockA{
%\IEEEauthorrefmark{1} School of Electronics Engineering and Computer Science, Peking University, Beijing, China.\\
%\IEEEauthorrefmark{2} School of Electrical and Information Engineering, The University of Sydney, Sydney, Australia.\\

%}}

\author{
\IEEEauthorblockN{\normalsize{Shuhang Zhang$^{1}$, Boya Di$^{1}$, Lingyang Song$^{1}$, {\em{Senior Member}}, {\emph{IEEE}}, \\ Yonghui Li$^2$, {\emph{Senior Member}}, {\emph{IEEE}} \\}}
\IEEEauthorblockA{$^1$\small{School of Electronics Engineering and Computer Science, Peking University, Beijing, China.} \\
$^2$\small{School of Electrical and Information Engineering, The University of Sydney, Sydney, Australia.} \\
\small{Email: \{zhangshuhang, diboya, lingyang.song\}@pku.edu.cn, yonghui.li@sydney.edu.au}
}
}
\maketitle
\setlength{\abovecaptionskip}{0pt}
\setlength{\belowcaptionskip}{-10pt}
%\titlespacing{\section} {0pt}{9pt}{0pt}
\begin{abstract}
In this paper, we study the resource allocation problem for a single-cell non-orthogonal multiple access (NOMA) relay network where an OFDM amplify-and-forward (AF) relay allocates the spectrum and power resources to the source-destination (SD) pairs. We aim to optimize the resource allocation to maximize the average sum-rate. The optimal approach requires an exhaustive search, leading to an NP-hard problem. To solve this problem, we propose two efficient many-to-many two-sided SD pair-subchannel matching algorithms in which the SD pairs and sub-channels are considered as two sets of players chasing their own interests. The proposed algorithms can provide a sub-optimal solution to this resource allocation problem in affordable time. Both the static matching algorithm and dynamic matching algorithm converge to a pair-wise stable matching after a limited number of iterations. Simulation results show that the capacity of both proposed algorithms in the NOMA scheme significantly outperforms the conventional orthogonal multiple access scheme. The proposed matching algorithms in NOMA scheme also achieve a better user-fairness performance than the conventional orthogonal multiple access.
\end{abstract}

\begin{IEEEkeywords}
Non-orthogonal multiple access, successive interference cancellation, relay network, resource allocation, matching theory,
proportional fair.

\vspace{2cm}
\hspace{-1cm}
Part of the material in this paper was presented in IEEE ICC, Kuala Lumpur, Malaysia, May 2016 \cite{ZDSL2016}.
\end{IEEEkeywords}

\section{Introduction}
With rapidly increasing demands in mobile services, wireless networks require an ever higher spectral efficiency and massive connectivity \cite{DWYHIW2015}. However, the capacity of conventional orthogonal frequency division multiple access (OFDMA) is not likely to afford the explosive growth of data traffic. As a result, finding new multiple access techniques to achieve high spectrum efficiency and massive connectivity have become a critical and urgent challenge to be resolved in the current wireless communication networks \cite{TGWIJFA2014}. The non-orthogonal multiple access (NOMA) technique was discussed to be used in LTE network in \cite{YYYHLLT2016}, and is regarded as a promising technology for 5G network \cite{HB2015}. Unlike the OFDMA scheme, NOMA can accommodate multiple users in the same time and frequency domains by differentiating the users through power domain or code domain multiplexing. NOMA has the advantages of a low complexity receiver and high spectrum efficiency due to its multiplexing nature.

In the conventional OFDMA scheme, each sub-channel can only be assigned to one user. In contrast, NOMA system allows multiple users to share the same sub-channel to achieve multiplexing gains. However, this also leads to unavoidable co-channel interference. To tackle this problem, various multi-user detection (MUD) techniques have been proposed, such as the successive interference cancellation (SIC) techniques~\cite{LMZ2014}, which can be applied at the end-user receivers to decode the received signals and reduce the inter-user interference effectively. In~\cite{YHBLLJ2015}, NOMA transmitter and low complexity receiver were proposed and its performance was compared with the theoretical performance of SIC. In \cite{DBSL2015}, the resource allocation and user scheduling problem was studied in a downlink NOMA network with a joint algorithm.

The relaying technology has been regarded as an effective method to extend the coverage and improve the system performance of wireless networks. By accommodating more users, we utilise NOMA in a relay network, where different users share the spectrum resource of the network. NOMA relay network has the advantage of providing massive connectivity and higher spectrum efficiency for the users. It also provides a larger power and coverage with the relay to improve the network performance. However, the design of NOMA relay is very challenging due to the complicating characteristics of relay networks.

Recently, there are some initial works on NOMA relay networks. In \cite{MG2015}, the outage probability of an amplify-and-forward (AF) relay network was derived and a lower bound of the outage probability was provided. In \cite{MVB2015}, joint network channel coding and decoding for half-duplex multiple access multiple relay channels in NOMA scheme was studied. The application of simultaneous wireless information and power transfer in NOMA network was studied in \cite{LDEP2016} and a cooperative protocol was developed. Cooperative NOMA is a technique that improves the quality of service of the network, where different users in a NOMA network cooperate with each other to enhance the performance. In \cite{ZMXDF2016}, a full-duplex device-to-device aided cooperative NOMA scheme was proposed to improve the outage performance of the weak users. A theoretical study on the selection of the cooperative NOMA was studied in \cite{DDP2016}. Although there are some works in NOMA relay networks, few of them have considered the resource allocation problem in such a network. Most of the existing works of NOMA relay network \cite{MG2015,MVB2015,LDEP2016} focused on the performance analysis, such as outage probability \cite{MG2015}, or focus on code design \cite{MVB2015}.
%Among these scenarios, resource allocation is undoubtedly one of the most important ones. A more efficient manner in resource
%allocation is attracting to the researchers since NOMA can take advantage of the multiplexing sufficiently. A novel adaptive
%power and frequency resource allocation algorithm in the NOMA system is proposed for performance enhancement in \cite{LBLH2014}
%to improve the cell average throughput. In \cite{KL2015}, the capacity of a basic decode-and-forward (DF) relay network was
%studied, and a sub-optimal power allocation scheme for NOMA was given. In \cite{DFP2015}, user pairing with fixed power allocation and cognitive radio inspired NOMA are studied to enhance the efficiency of network resource. In \cite{DBSL2015}, the author studied resource allocation and user scheduling problem for a downlink non-orthogonal multiple access network where the base station allocates spectrum and power resources to a set of users with a joint algorithm.

In this paper, we consider the NOMA relay networks to enhance the access spectral efficiency and at the same time provide wide area coverage in a large scale network. An OFDM amplify-and-forward (AF) relay assigns the sub-channels and allocate different level of power to a set of source destination (SD) pairs, each of which consists of a source node and a destination node, the source node transmits through the relay to its paired destination node. Each SD pair can occupy multiple sub-channels and each sub-channel can be shared by multiple SD pairs. For the SD pairs sharing the same sub-channel, the SIC technique is adopted to remove the inter-user interference. Joint sub-channel and power allocation is then formulated as a non-convex optimization problem to maximize the total sum-rate. The optimal solution for this problem is NP-hard and requires exhaustive search. Therefore, an efficient low complexity resource allocation algorithm is required. The process of solving resource allocation problem can be associated with matching theory due to the structure of the system. In a NOMA relay network, the relay allocates a set of spectrum resources to the set of SD pairs. The allocation process can be solved with matching theory under NOMA protocol, and the design of matching algorithm plays an important part in the resource allocation problem.

We propose to utilize matching theory in the resource allocation problem in NOMA relay network. This problem is separated into two subproblems, a sub-channel allocation problem and a power allocation problem. In the sub-channel allocation problem, the set of SD pairs and the set of sub-channels are both seeking to match with the opposite set to maximize their own profit. Therefore, we consider the SD pairs and sub-channels as two sets of selfish and rational players aiming at maximizing their own profits. Motivated by this, we formulate the sub-channel allocation problem as a many-to-many two sided matching game with externalities in which interdependencies exist between the players' preferences due to the co-channel interference. Two novel user-spectrum matching algorithms extended from the Gale-Shapley algorithm \cite{RS1992} are proposed for the matching game formulation to reach a stable matching. For the power allocation problem, the water filling algorithm is utilized to enhance the power efficiency. We take proportional fairness into consideration and aim at maximizing a function related to the average throughput of all the users, so as to guarantee the throughput of the users at the edge of the cell with a poor channel condition.

The main contributions of this paper can be summarized as follows.
\begin{enumerate}[(1)]
\item We formulate a joint sub-channel and power allocation problem for a downlink NOMA network to maximize the average sum-rate over each sub-channel.
\item We represent the sub-channel allocation problem equivalent to a many-to-many matching game, and propose two matching algorithms considering externalities \cite{PY2015}, that is, the change in the matching structure caused by inter-user interference is fully embodied. We then utilize an iterative water filling algorithm \cite{GHLP2009} to allocate the power. The properties of our matching algorithm are then analyzed in terms of stability, convergence and complexity.
\item Simulation results show that the proposed matching algorithms in NOMA scheme outperform OFDMA scheme significantly in both capacity and fairness.
\end{enumerate}

The rest of this paper is organized as follows. In Section \uppercase\expandafter{\romannumeral2} we describe the system model
of the NOMA relay networks. In Section \uppercase\expandafter{\romannumeral3}, we formulate the optimization resource allocation
problem as a many-to-many two-sided matching problem, and propose a matching algorithm, followed by the corresponding analysis.
Simulation results are presented in Section \uppercase\expandafter{\romannumeral4}, and finally we conclude the paper in
Section \uppercase\expandafter{\romannumeral5}.

\section{System Model}
We consider a single-cell one-way NOMA network as depicted in Fig.1, consisting of one OFDM AF relay $R$ and $N$ SD pairs. Each SD pair consists of one source node and one destination node, where the source node communicates with the destination node assisted by relay $R$. Let ${\cal{S}} = \{ 1,2, \cdots N\}$ denote the set of source nodes and ${\cal{D}} = \{ 1,2, \cdots N\}$ denote the set of destination nodes. We assume that relay $R$ has full knowledge of the instantaneous channel side information (CSI), which varies in every time slot. In each time slot, based on the CSI of each channel, relay $R$ assigns a subset of non-orthogonal sub-channels, denoted as ${\cal{K}} = \{ 1,2, \cdots K\}$, to the SD pairs and allocates different power over
the sub-channels. According to the NOMA protocol\cite{SKBNLH2013}, one sub-channel can be allocated to multiple SD pairs, one SD
pair has access to multiple sub-channels in the network, and each SD pair shares the same group of sub-channels. We assume that at most $q_{u}$ SD pairs can have access to each sub-channel. To guarantee fairness among the SD pairs, we set a maximum number of $q_{l}$ sub-channels allocated to each SD pair~\cite{BLHVL-2014}. Communication between the source nodes and destination nodes in each time slot consists of two phases, described specifically as follows.%\vspace{-3mm}
\begin{figure}[!ht]
\centerline{\includegraphics[width=12cm]{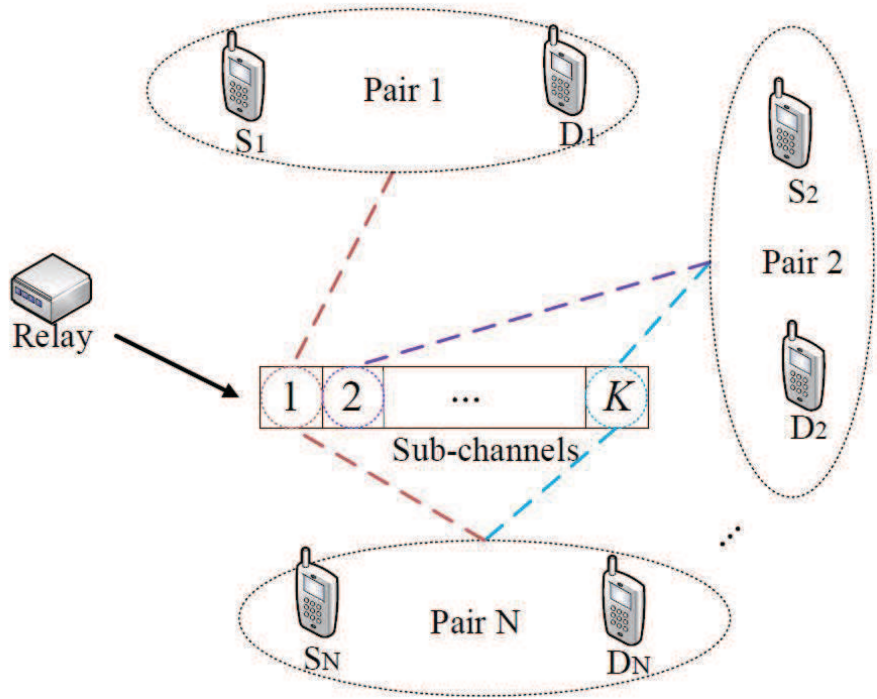}}
%\vspace{-5mm}
\caption{System model of the single-cell one-way NOMA network.}
%\vspace{-2mm}
\end{figure}

In the first phase, the source nodes transmit signals to relay $R$. We denote the $m$th source node as $S_{m}$ and the $k$th
sub-channel as $SC_{k}$. The transmitting power of $S_{m}$ over $SC_{k}$ is denoted as p$_{k,m}$, satisfying
$\sum_{k=1}^{K} p_{k,m} \leq {P_{SN}}$ for each $S_{m}\in\cal{S}$, where P$_{SN}$ is the maximum transmit power of $S_{m}$. We consider a block fading channel, for which the channel remains constant within each time-slot, but varies independently from one to another. The complex coefficient of $SC_{k}$ between $S_{m}$  and relay $R$ consists of two parts, the small-scale fading and the large-scale fading \cite{NAYLM2016}\cite{LJJZM2016}. It is denoted by $h_{k,m}=g_{k,m} {/} \left({d_{m}}\right)^{\alpha}$, where $g_{k,m}$ describes the Rayleigh fading channel gain of $SC_{k}$ from $S_{m}$ to relay $R$, ${d_{m}}$ describes the distance between relay $R$ and $S_{m}$, with $\alpha$ being the constant path loss coefficient. The Rayleigh fading channel gain $g_{k,m}$ is a small-scale fading, whose real part and imaginary part of the channel gain both obey the Gaussian distribution, and it varies in different time slots. The large-scale fading is denoted as $1 {/} \left({d_{m}}\right)^{\alpha}$, which only depends on the distance between the source node and relay $R$ and remain constant in different time slots. Let $x_{k,m}$ be the transmitting information symbol of unit energy from $S_{m}$ over $SC_{k}$. The signal that relay $R$ receives from $S_{m}$ over $SC_{k}$ is given by
\begin{eqnarray}
z_{k,m}=h_{k,m}\sqrt{p_{k,m}}x_{k,m}+n_{k},\label{system_4}
\end{eqnarray}
where $n_{k}\sim \mathcal {N}{\left( {0,\sigma_{s}^{2}}\right)}$ is the additive white Gaussian noise (AWGN), and
$\sigma_{s}^{2}$ is the noise variance.

In the second phase, relay $R$ amplifies the signals received from each source node and broadcasts the superposed signals to
the destination nodes \cite{MG2015}. Let $G_{k}$ denote the amplification factor of relay $R$ over $SC_{k}$ and $q_{k,m}$
is the transmit power that relay $R$ allocates to $D_{m}$ over $SC_{k}$, where $D_{m}$ is the $m$th destination node. The
relation between $q_{k,m}$ and $G_{k}$ is given by
\begin{eqnarray}
q_{k,m}=G_{k}^{2}\left(p_{k,m}\left|h_{k,m}\right|^{2}+\sigma_{s}^{2}\right).
\end{eqnarray}
Correspondingly, we denote $f_{k,m}$ as the complex coefficient of $SC_{k}$ between relay $R$ and $D_{m}$, and
$f_{k,m}=c_{k,m}{/}\left({b_{m}}\right)^{\alpha}$, where $c_{k,m}$ denotes the Rayleigh fading channel gain of $SC_{k}$ from
relay $R$ to $D_{m}$, and $b_{m}$ is the distance between $D_{m}$ and relay $R$. The small-scale fading channel gain $c_{k,m}$ differs in every time slot, while the large-scale fading channel gain $1 {/} \left({b_{m}}\right)^{\alpha}$ remain constant. Let $U_{k}$ be the set of SD pairs that have access to $SC_{k}$. The signal that $D_m$ receives from relay $R$ over $SC_k$ is given by
\begin{eqnarray}
y_{k,m}=G_{k}f_{k,m}\sum_{i\in{U_{k}}}z_{k,i}+w_{k},\label{system_11}
\end{eqnarray}
where $w_{k}\sim \mathcal {N}{\left( {0,\sigma_{d}^{2}}\right)}$ is AWGN. It is assumed that $\sigma_{d}^2=\sigma_{s}^2$. By
substituting $\left( \ref{system_4} \right)$ into $(\ref{system_11})$, the equation can be rewritten as
\begin{eqnarray}
y_{k,m}=&G_{k}f_{k,m}\left(\sum_{i\in{U_{k}}}{h_{k,i}\sqrt{p_{k,i}}x_{k,i}+n_{k}}\right)+w_{k}\\
=&G_{k}f_{k,m}h_{k,m}\sqrt{p_{k,m}}x_{k,m}+G_{k}f_{k,m}\sum_{i\in{U_{k}},i\neq m}h_{k,i}\sqrt{p_{k,i}}x_{k,i}+G_{k}f_{k,m}n_{k}+w_{k}.\label{system_16}
\end{eqnarray}

From $\left( \ref{system_16} \right)$, we see that the equivalent channel gain of the a SD pair is determined by the channel gains of both the first and the second phases. We then normalize $\left( \ref{system_16} \right)$ and extract
the equivalent channel gain as
\begin{eqnarray}
\gamma_{k,m}=\frac{{G_{k}}^2{\left|f_{k,m}\right|^{2}}{p_{k,m}}{\left|h_{k,m}\right|^2}}{{G_{k}}^2{\left|f_{k,m}\right|}^2{n_{k}}^2+{w_{k}}^2}.
\end{eqnarray}
The numerator part shows the channel fading and power of the target signal, and the denominator part describes the AWGN of both source-relay transmission and relay-destination transmission. After receiving the signals, the destination nodes perform SIC to reduce the interference from the source nodes of other SD pairs with a smaller equivalent channel gain over $SC_{k}$ \cite{UKH-2012}. For example, for $S_{i},S_{j}\in U_{k}$, if $\left|\gamma_{k,i}\right|^{2}>\left|\gamma_{k,j}\right|^{2}$, $D_{i}$ first treats $S_j$ as the interference to $S_i$ and cancel $S_j$ in decoding $S_i$\footnote{For the downlink channel that we consider, \cite{T2005} shows that a user can decode the signal of another user with worse channel gain, with any split of the total power.}. The order for decoding is based on the increasing channel gains described above, which guarantees that the upper bound on the capacity region can be reached\cite{WCYW2014}. The interference that $D_{m}$ receives over $SC_{k}$ is shown as below,
%From $\left( \ref{system_16} \right)$, we see that the channel gain of the relay network is complex in expression. Although it is determined by both the channel gain of the second phase and the channel gain of the first phase, $f_{k,m}$ has a much larger impact than $h_{k,m}$ since $h_{k,m}$ and all $h_{k,i} \left(i\in{U_{k}}\right)$ affect the channel gain of $y_{k,m}$ together. For simplicity, we only consider the channel gain of the second phase, i.e., $f_{k,m}, k\in{\cal{K}}, m\in{\cal{D}}$. For example, for $S_{i},S_{j}\in U_{k}$, if $\left|f_{k,i}\right|^{2}>\left|f_{k,j}\right|^{2}$, $D_{i}$ can cancellate the signal-to-signal interference of $S_{j}$ from $S_{i}$ while decoding. The order for decoding is based on the increasing channel gains described above, which guarantees that the upper bound on the capacity region can be reached\cite{WCYW2014}. The interference that $D_{m}$ receives over SC$_{k}$ is shown as below,
\begin{eqnarray}
I_{k,m}=\sum\nolimits_{\left|\gamma_{k,i}\right|^{2}>\left|\gamma_{k,m}\right|^{2}}{G_{k}^{2}{\left|{f_{k,i}}\right|^{2}}
{p_{k,m}}{\left|{h_{k,i}}\right|^{2}}}.
\end{eqnarray}
Note that the noise and interference for $D_{m}$ over $SC_{k}$ consists of three parts: the noise at $D_{m}$, the amplified noise
forwarded by relay $R$, and interference from other source nodes. Therefore, the data rate of $D_{m}$ over $SC_{k}$ is given by
\begin{eqnarray}
R_{k,m}=\log_2\left({1+\frac{G_{k}^{2}{\left|{f_{k,m}}\right|^{2}}{p_{k,m}}{\left|{h_{k,m}}\right|^{2}}}{\sigma_{d} ^{2}+G_{k}^{2}{\left|{f_{k,m}}\right|^{2}}\sigma_{s}^{2}+I_{k,m}}}\right).\label{system_0}
\end{eqnarray}

When considering a multi-time slot scenario, we care more about the average throughput than the instantaneous throughput of the network. The average throughput that $D_{m}$ received per sub-channel in the $\left(t+1\right)$th time slot is defined \cite{OKH2012} as
\begin{eqnarray}
T_{m}(t+1)=\left(1-\frac{1}{t_c}\right)T_{m}(t)+\frac{1}{t_c}\left(\frac{1}{K}\sum_{k=1}^{K}R_{k,m}(t)\right).
\end{eqnarray}
Parameter $t_c$ denotes the time duration during which we calculate the average throughput for throughput averaging, and $R_{k,m}(t)$ is the data rate of $D_{m}$ over $SC_{k}$ in the $t$th time slot, which can be calculated by~$\left( \ref{system_0} \right)$.

We utilize proportional fairness in this model to guarantee the quality of service for the cell edge users. The scheduling metric of the SD pairs over sub-channel $SC_k$ is shown as follows~\cite{KG2005},
\begin{eqnarray}
{F_k}(D) = \prod\limits_{m \in D} {(1 + \frac{{{R_{k,m}}(t)}}{{({t_c} - 1){T_m}(t)}})}.
\end{eqnarray}
The scheduling metric for the set of SD pair is $F_k(\cal{D})$, and the SD-subchannel pairing that maximizes equation (10) for the destination nodes in $\cal{D}$ over each sub-channel will be adopted.

Our objective is to maximize ${F_k}(D)$ over each sub-channel by jointly allocating $\left\{{{\phi _{m,k}}}, p_{m,k}, q_{m,k}\right\}$ in each time slot. To better describe the resource allocation, we define a binary $N \times K$ SD pair-subchannel pairing matrix $\bm{\Phi}$, in which $\phi_{m,k}=1$ denotes that S$_{m}$ and D$_{m}$ are paired with $SC_{k}$. We also assume that relay $R$ has a maximum transmitted power of $Q_{R}$, $q_{k,m}$ and $Q_{R}$ satisfy the following inequation $\sum_{k\in{\cal{K}}}\sum_{m\in{\cal{D}}} q_{k,m} \leq {Q_{R}}$. The objective and restrictions of problem can be formulated as follows:
\begin{subequations} \label{system_optimization}
\begin{align}
&\mathop{\max }\limits_{{\textbf{$\bm{\Phi,p,q}$}}}\sum_{k=1}^{K}{F_k(\cal{D})},\label{system_1}\\
&\textbf{\emph{s.t. }}{\sum_{m=1}^{N}{\phi_{m,k}\leq q_{u}}},{\forall k \in {\cal K}},\label{system_2}\\
&\quad\, \, \,{\sum_{k=1}^{K}{\phi_{m,k}\leq q_{l}}},{\forall m \in {\cal S}},\label{system_17}\\
&\quad\, \, \,{\sum_{k=1}^{K} p_{k,m} \leq {P_{SN}}}, \forall m \in {\cal{S}},\label{system_5}\\
&\quad\, \, \,{\sum_{k\in{\cal{K}}}\sum_{m\in{\cal{D}}} q_{k,m}\phi_{k,m} \leq {Q_{R}}},\label{system_3}\\
&\quad\, \, \,{p_{k,m} \geq 0},{\forall k \in {\cal K}},{\forall m \in {\cal S}},\label{system_6}\\
&\quad\, \, \,{q_{k,m} \geq 0},{\forall k \in {\cal K}},{\forall m \in {\cal S}},\label{system_7}\\
&\quad\, \, \,\phi_{k,m}\in{\left\{0,1\right\}},\label{system_8}
\end{align}
\end{subequations}
where $\left( \ref{system_2} \right)$ and $\left( \ref{system_17} \right)$ shows that each sub-channel can be allocated to at
most $q_{u}$ SD pairs and each SD pair can have access to at most $q_{l}$ sub-channels. $\left( \ref{system_5} \right)$ and
$\left( \ref{system_3} \right)$ are the power restrictions of the source nodes and relay $R$ respectively. Constraints
$\left( \ref{system_6} \right)$ and $\left( \ref{system_7} \right)$ show that the transmitting power is no less than 0.

It can be observed that $\left( \ref{system_1} \right)$ is a non-convex problem due to the binary constraint in
$\left( \ref{system_8} \right)$ and the existence of the interference term in the objective function \cite{WN2014}. However, in
the complexity theory, non-convexity can not prove the problem¡¯s hardness, as a problem could be inappropriately formulated.
Therefore, we prove that the non-convex optimization problem $\left( \ref{system_1} \right)$ is also an NP-hard one in the
following theorem.

\textbf{Theorem 1:} The sum-rate maximization problem in $\left( \ref{system_1} \right)$ is NP-hard.

\begin{proof}
See Appendix A.
\end{proof}
\section{Many to Many Matching for NOMA}
As shown in $\left( \ref{system_1} \right)$, the average sum-rate of the network is determined by both power allocation and
sub-channel allocation. We decouple the resource allocation problem into two subproblems, sub-channel allocation and power
allocation and develop a sub-optimal solution.

There are some previous works on resource allocation problems that have utilised matching theory. A distributed spectrum access algorithm for cognitive radio relay networks that results in a stable matching was proposed in \cite{BLVL2013} and \cite{BLLV2011}. In \cite{LJZH2011}, a marketing noncooperation game was applied in proposing a dynamic spectrum sharing algorithm to reach the Nash equilibrium. In \cite{HH2015}, the authors focused on relay-aided D2D communication, and proposed a distributed solution approach using stable matching to allocate radio resources. In the sub-channel allocation, we assume that each source node allocates its power equally over the sub-channels, and the amplification coefficient of relay $R$ over each sub-channel are the same. We recognize sub-channel allocation as equivalent to a many-to-many two-sided matching problem between the set of SD pairs and the set of sub-channels, which we will explain in details in Section III.A. After the sub-channel allocation is performed, each source node allocates its own transmitting power over its matched sub-channels by utilizing water filling algorithm~\cite{YRBC2004}, then relay $R$ determines its amplification coefficient over each sub-channel.

\subsection{Two Sided Many-to-Many Matching Problem Formulation}
In the sub-channel allocation process, the SD pairs prefers to access the sub-channels with good quality to achieve the best service, while relay $R$ aims at maximizing the throughput of the network by arranging which SD pairs can be assigned to each sub-channel. The sub-channel allocation problem can be considered by relay $R$ as a matching process in which the set of sub-channels and the set of SD pairs match with each other. To better describe the matching process between the SD pairs and the sub-channels\footnote{The outcome of the matching is the solution for relay $R$ to allocate the sub-channels as the interests of the sub-channels and relay $R$ are identical.}, we consider the set of SD pairs and the set of sub-channels as two disjoint sets of selfish and rational players aiming to maximize their own interests. Since the source and destination nodes are already paired, we can transform the matching problem between the set of SD pairs and the set of sub-channels into an equivalent matching process between source nodes $\cal{S}$ and sub-channels $\cal{K}$ for convenience.

Let $(S_m, SC_k)_t$ denote a matching pair if $SC_{k}$ is assigned to $S_{m}$ in the $t$th time slot. To depict the influence of the resource allocation for each player, we assume that each player has preferences over the subsets of the opposite set. The preference of each source node is based on its achievable data rates, while the preference of the sub-channel $SC_k$ is determined by $F_k(\cal{D})$. Note that the sub-channel allocation is performed in each time slot according to the corresponding CSI, implying that the matching process can be considered separately for each time slot. Without loss of generality, we consider the matching process in time slot $t$. We define the \emph{utility} of sub-channel $SC_k$ as the product of the source nodes¡¯ average throughput over $SC_k$. Given $V$ and $V'$ as two subsets of source node $S$, $SC_k$'s preference over different subsets of source nodes can be written as
\begin{eqnarray}
\begin{split}
{V}\succ{}_{SC_{k}}{V^{'}}, {V}\subseteq {S}, V^{'} \subseteq {S} \Leftrightarrow F_k({V}) > F_k({V'}),
\end{split}
\end{eqnarray}
which implies that $SC_{k}$ prefers $V$ to ${V{'}}$ because the former subset of source nodes provides a larger utility than the latter one. Given $U$ and $U'$ as two subsets of sub-channel $\cal{K}$, the preference of $S_{m}$ over these subsets of sub-channels can be represented as
\begin{eqnarray}
\begin{split}
{U}\succ{}_{S_{m}}{U^{'}}, {U}\subseteq {K}, &{U^{'}} \subseteq {K} \Leftrightarrow \\
\sum_{t\in U}\frac{{G_0^2}{p_{0}}\left|f_{t,m}\right|^{2}\left|h_{t,m}\right|^{2}}{\sigma_{d}^{2}+{G_0^2}\left|f_{t,m}\right|^{2}\sigma_{s}^{2}+I_{t,m}}>&
\sum_{t'\in U'}\frac{{G_0^2}p_{0}\left|f_{t{'},m}\right|^{2}\left|h_{t{'},m}\right|^{2}}{\sigma_{d}^{2}+{G_0^2}\left|f_{t{'},m}\right|^{2}\sigma_{s}^{2}+I_{t',m}},
\label{system_13}
\end{split}
\end{eqnarray}
where $G_0$ and $p_0$ are the amplification coefficient of relay $R$ and the transmitting power of each source node. Inequality $\left( \ref{system_13} \right)$ implies that $S_m$ prefers $U$ to $U'$ due to a higher channel gain.

\textbf{Definition 1:} A preference list is an ordered set containing all the possible subsets of the opposite set for player
$i$ ($i \in \cal{K} \cup \cal{S}$). Given $A_1, A_2, \cdots, A_n$ are the subsets of the opposite set of player $i$, player $i$'s preference list $P\left(i\right)=\{ A_1, A_2, \cdots, A_n\}$ representing that  $A_1, A_2, \cdots, A_n$ are player $i$'s potential matching pairs and ${A_1 \succ_i A_2 \succ_i \cdots \succ_i A_n}$.

We denote $\textbf{\textit{P}}=\left\{P\left(S_{1}\right), P\left(S_{2}\right) \cdots, P\left(S_{N}\right), P\left(SC_{1}\right), P\left(SC_{2}\right) \cdots, P\left(SC_{K}\right)\right\}$ as the set of preference lists of the source nodes and sub-channels, where $P\left(S_{m}\right)$ and $P\left(SC_{k}\right)$ are the preference lists of $S_{m}$ and $SC_{k}$, respectively. We also assume that the preferences of the source nodes and sub-channels are \emph{transitive}. The definition of \emph{transitive} is shown as follows.

\textbf{Definition 2:} We say the preference of $m$ is \emph{transitive} if for  ${L}\succ{}_{m} {L'}$ and $ {L'}\succ{}_{m} {L''}$, we have $L\succ{}_{m} {L''}$, where $m$ is a player of the matching game. $L$, $L'$ and $L''$ are the subsets of player $m$'s opposite set.

With the definition of \emph{transitive} and preference list, we can then formulate the optimization problem as a many-to-many
two-sided matching game.% \cite{HSD2014}.

\textbf{Definition 3:} Given two disjoint sets, $\cal{S}= \{ \text{1},\text{2}, \cdots, \text{S}\}$ of the source nodes, and
$\cal{K} = \{\text{1}, \text{2}, \cdots, \text{K}\}$ of the sub-channels, a many-to-many matching $\bm{\Psi}^t$ is a mapping from
the set $\cal{S}\cup \cal{K}$ into the subsets of $\cal{S}\cup \cal{K}$ in the $t$th time slot such that for every
$S_{m} \in \cal{S}$, and $SC_{k} \in \cal{K}$:
\begin{enumerate}[1)]
\item $\bm{\Psi}^t\left(S_{m}\right) \subseteq \cal{K},\label{system_9}$
\item $\bm{\Psi}^t\left(SC_{k}\right) \subseteq \cal{S},\label{system_10}$
\item $\bm{\Psi}^t\left|\left(SC_{k}\right)\right| \leq q_{u},\label{system_12}$
\item $\bm{\Psi}^t\left|\left(S_{m}\right)\right| \leq q_{l},\label{system_15}$
\item $SC_{k}\in \bm{\Psi}^t\left(S_{m}\right) \Leftrightarrow S_{m}\in \bm{\Psi}^t\left(SC_{k}\right).\label{system_14}$
\end{enumerate}
Conditions $\ref{system_9})$ and $\ref{system_10})$ state that each source node is matched with a subset of sub-channels, and each sub-channel can be allocated to a subset of source nodes. Condition $\ref{system_12})$ and $\ref{system_15})$ implies that each sub-channel can be allocated to no more than $q_{u}$ SD pairs, and each SD pair can be allocated to no more than $q_{l}$
sub-channels. Condition $\ref{system_14})$ shows that the sub-channels and SD pairs are matched mutually in each time slot. We define the \emph{strategy} of player $i$ as the set of the players that are matched with $i$, i.e., $\bm{\Psi}^t\left(i\right)$. A player's \emph{strategy} is obtained from the many-to-many matching process, based on the preference of each player and the matching algorithm.

The matching model above is more complicated than the conventional two-sided matching models for two main reasons. Firstly, for
source node $S_i$, its strategy is not only determined by itself, but also affected by the strategies of other source nodes, which is called \emph{externalities}. Because of the externalities, the preference list of the source node varies with the matching structure\footnote{Matching structure describes the current matching condition between the set of source nodes and the set of sub-channels.}. For sub-channel $SC_k$, the utility it can obtain from the player of the opposite set is affected by $\bm{\Psi}^t\left(SC_{k}\right)$ because of the co-channel interference. As a result, in our
model the players should be matched with any subsets of the opposite set instead of a single player. Therefore, the
number of potential matching combinations can be extremely huge with the increment of the players in each set. This makes the
problem quite intractable even when the power allocation is not considered. Secondly, under the conventional definition of stable
matching such as that in \cite{RS1992}, there is no guarantee that a stable matching exists even in many-to-one matchings. In
that case, the matching is possible to be unstable in this matching model. For these reasons, the matching process of this
model is sophisticated and there is no existing matching algorithm that can solve this problem efficiently. Therefore, to solve this matching problem, we develop two extended versions of the Gale-Shapley algorithm~\cite{RS1992} and propose two new matching
algorithms in Section \uppercase\expandafter{\romannumeral3}.B and Section \uppercase\expandafter{\romannumeral3}.C.

\subsection{Static SD pair-Subchannel Matching Algorithm}

In this subsection, we propose a low-complexity static SD pair-subchannel matching algorithm (SSD-SMA).
To achieve the low complexity and operability of this algorithm, we assume that all the source
nodes only construct their preference lists at the beginning of the matching. The preference lists of the
source nodes are static and will not be changed throughout the matching process, which is the same as
the conventional Gale-Shapley algorithm. In this matching model, every source node makes their decisions
first according to their own preference lists. In each round of proposals from source nodes, each source
node proposes itself to at most one sub-channel and then wait the response from the sub-channels. After all
the incompletely matched\footnote{A source node is incompletely matched when it is matched with less than $q_l$ sub-channels.} source nodes have proposed themselves to the sub-channels, the sub-channels
decide whether to accept the proposing source nodes. We define it as a \emph{static matching iteration} when the
source nodes propose themselves to the set of sub-channels and at least one proposal is accepted.

It is also assumed that each source node has no idea of other source nodes' preference when constructing
their preference lists. That is, for each source node $S_i$, the potential externalities brought by other
source nodes are unpredictable at the beginning of the matching. As a result, the source nodes do not
consider the impact of potential co-channel interference brought by other source nodes when constructing
preference lists. We can then simplify the preference list of the source nodes in SSD-SMA in
the following method. The subsets of $K$ can be replaced by the sub-channels in $K$ in the preference lists
of the source nodes. Given $k$ and $k'$ as two different sub-channels, $\left(\ref{system_13}\right)$ can be simplified as
\begin{eqnarray}
\begin{split}
{k}\succ{}_{S_{m}}{k^{'}}, {k}\subseteq {K}, &{k^{'}} \subseteq {K} \Leftrightarrow \\
\frac{{G_0^2}p_{0}\left|f_{k,m}\right|^{2}\left|h_{k,m}\right|^{2}}{\sigma_{d}^{2}+{G_0^2}\left|f_{k,m}\right|^{2}\sigma_{s}^{2}}>&
\frac{{G_0^2}p_{0}\left|f_{k{'},m}\right|^{2}\left|h_{k{'},m}\right|^{2}}{\sigma_{d}^{2}+{G_0^2}\left|f_{k{'},m}\right|^{2}\sigma_{s}^{2}}.
\end{split}
\end{eqnarray}
We then define the static preference list of the source nodes.

\textbf{Definition 4:} A static preference list of a source node is an ordered set containing all the possible sub-channels for
source node $S_i$ ($S_i \in \cal{S}$). Given $SC_{B_1}, SC_{B_2}, \cdots, SC_{B_n}$ as the sub-channels to which $S_i$ is possible to access with, $S_i$'s static preference list $P_s\left(S_i\right)=\{SC_{B_1}, SC_{B_2}, \cdots, SC_{B_n}\}$ represents that $SC_{B_1}, SC_{B_2}, \cdots, SC_{B_n}$ are $S_i$'s potential matching pairs and ${SC_{B_1} \succ_{S_{i}} SC_{B_2} \succ_{S_{i}} \cdots \succ_{S_{i}} SC_{B_n}}$.

Let's denote by $\textbf{\textit{P}}_s=\left\{P_s\left(S_{1}\right), P_s\left(S_{2}\right) \cdots, P_s\left(S_{N}\right)\right\}$ the set of static preference lists of the source nodes, where $P_s\left(S_{m}\right)$ is the preference lists of $S_{m}$. However, the preference lists of the sub-channels are unreducible in static matching iterations because there still exist interdependencies between the source nodes who share the same sub-channel in SSD-SMA. The key idea of SSD-SMA is that each source node proposes itself to the most preferred sub-channel which has not refused it. The sub-channels then decide whether to accept these proposals by judging if it can bring
benefits to the itself. When all the proposed sub-channels have responded to the proposing source
nodes, this static matching iteration is performed and the source nodes will check if it is necessary to
perform the next static matching iteration. We now describe how the sub-channels choose the proposing
source nodes by introducing the concept of $\emph{blocking pair}$.

$\textbf{Definition 5:}$ Given a matching $\bm{\Psi}^t$ and a pair $\left(S_{m}, SC_{k}\right)_t$ with
$S_{m}\notin\bm{\Psi}^t(SC_{k}$) and $SC_{k}\notin\bm{\Psi}^t (S_{m})$. $(S_{m}$, $SC_{k})_t$ is a \emph{blocking pair} if
$\left(1\right) \left\{SC_{k}\right\} \in P_s\left(S_{m}\right)$. $\left(2\right)$
$\left\{S_{m}\right\}\cup\bm{\Psi}^t(SC_{k})\succ {}_{SC_{k}}\bm{\Psi}^t(SC_{k})$. $\left(3\right)SC_{k}\cup\bm{\Psi}^t(S_{m})\succ{}_{S_{m}}\bm{\Psi}^t(S_{m})$.

The existence of blocking pair $\left(S_{m}, SC_{k}\right)_t$ has the following necessary conditions. Firstly, $S_{m}$ and $SC_{k}$ have never been matched with each other. That is to say, $S_{m}$ have not proposed itself to $SC_{k}$ before, and $SC_{k}$ is still in $S_{m}$'s preference list. Secondly, the matching of $SC_{k}$ and $S_{m}$ can increase both of their utility. A matching $\bm{\Psi}^t$ is blocked by $\left(S_{m}, SC_{k}\right)_t$ when both side of the players prefers to be matched with each other.

With the definition above, we can describe the strategy of each sub-channel as below. To reduce the complexity, the sub-channels do not construct the whole preference lists in advance, as will be analyzed in
Section \uppercase\expandafter{\romannumeral3}.E. When $SC_{j}$ receives the proposal from source node $S_{i}$,
$S_{i}$ and $SC_{j}$ form a blocking pair if $\{S_{i}\} \cup \bm{\Psi}^t(SC_{j})$ can provide a higher sum-rate than
$\bm{\Psi}^t(SC_{j})$ over $SC_{j}$. Under this condition, $SC_{j}$ will accept the proposal from
$S_{i}$. However, there is a special case when the sub-channel $SC_{j}$ has already been paired with $q_u$ source nodes before accepting the proposal from $S_{i}$. In that case, $SC_{j}$ has to give up one of the matched source nodes in the subset $\{S_{i}\} \cup \bm{\Psi}^t(SC_{j})$. The sub-channel $SC_{j}$ then calculates every possible $F_k(V)$ where $V \subseteq \{S_{i}\} \cup \bm{\Psi}^t(SC_{j})$ and $\left|V\right|=q_u$. Afterwards, $SC_{j}$ chooses to match with $\argmax\limits _V F_k\left(V\right)$. Suppose that $SC_{j}$ gives up the matching with $S_{m}$, $\left(S_{m}, SC_{j}\right)_t$ will not form a blocking pair any longer because $S_{m}$ will not propose itself to $SC_{j}$ again. The process of the proposed static SD pair-subchannel matching algorithm is to find and eliminate the potential blocking pairs.

We now describe the whole process of SSD-SMA. The specific details of the proposed SSD-SMA are described in Table \uppercase\expandafter{\romannumeral1}, consisting of an initialization phase and a matching phase.

In the initialization phase, each source node calculates the rate that every sub-channel can provide and then constructs its static preference list in the order of the corresponding rates. In the matching phase, each source node that has been paired with less than $q_l$ sub-channels proposes itself to the most preferred sub-channel in its static preference list if there is any, and remove the sub-channel from its static preference list. After all the proposing source nodes have proposed themselves to the set of sub-channel, the sub-channels that have received proposals will decide whether to accept the proposals of the source nodes or not. The sub-channel will accept the proposal from a source node if it can increase its throughput over itself by matching with this source node. If the sub-channel has already matched with $q_u$ source node before accepting the current proposal, it will unmatch with one of the matched source nodes that causes the minimum loss of throughput over it. After all the proposing source nodes have received the responses from the corresponding sub-channels, they will check if they are still willing to make any proposals. A source node will make a proposal when it is matched with less than $q_l$ sub-channels and still has a non-empty static preference list. If any source node wants to make a new proposal, another static matching iteration will be performed. The SSD-SMA terminates when no source node would like to make new proposals.

\begin{table}[!ht]
\renewcommand{\arraystretch}{2.0}
\caption{Static SD pair-subchannel matching algorithm (SSD-SMA)}
\centering
\begin{tabular}{p{160mm}}
\hline
\begin{enumerate}
\item \textbf{Initialization Phase}
    \begin{codebox}
    \zi \For i=1:N
    \zi \quad \For j=1:K
    \zi \quad\quad Calculate $R_{j,i}$;
    \zi \quad $S_{i}$ constructs $P_s\left(S_{i}\right), i\in \cal{S}$.\End\End
    \end{codebox}
\item \textbf{Matching Phase}
    \begin{codebox}
    \zi Static-matching-iteration=1;
    \zi Source-node-propose=0;
    \zi \While Static-matching-iteration==1 $\bm{or}$ Source-node-propose=1
    \zi \quad Static-matching-iteration+=1;
    \zi \quad\For i=1:N
    \zi \quad\quad\If $S_{i}$-matched-SC-number $\leq q_{l}$ $\bm{and}$ $P_s\left(S_{i}\right) \neq \varnothing$
    \zi \quad\quad\quad $S_{i}$ propose itself to $P_s\left(S_{i}\right)\left[1\right]$;
    \zi \quad\quad\quad Source-node-propose=1;
    \zi \quad\quad\quad Remove $P_s\left(S_{i}\right)\left[1\right]$ from $P_s\left(S_{i}\right)$;\End\End\End\End
    \zi \quad\For k=1:K
    \zi \quad\quad\If $SC_{k}$ has received any proposal (e.g.$S_{m}$)
    \zi \quad\quad\quad\If $\left\{S_{m}\right\}\cup\bm{\Psi}^t(SC_{k})\succ {}_{SC_{k}}\bm{\Psi}^t(SC_{k})$
    \zi \quad\quad\quad\quad $SC_{k}$ accept the proposal from $S_{m}$;
    %\zi \quad\quad\quad\quad $S_{m}$-matched-SC-number+=1;
    %\zi \quad\quad\quad\quad $SC_{k}$-matched-source-node-number+=1;
    \zi \quad\quad\quad\quad \If $SC_{k}$-matched-source-node-number $\ge q_u$
    \zi \quad\quad\quad\quad\quad \For j=1:N
    \zi \quad\quad\quad\quad\quad\quad \If $S_{j}$ and $SC_{k}$ are matched
    \zi \quad\quad\quad\quad\quad\quad\quad Calculate $F_k({\bm{\Psi}^t\left(SC_{k}\right)}\setminus S_{j})$
    \zi \quad\quad\quad\quad\quad Find the largest $F_k(D)$ in the circulation above (e.g.$F_k({\bm{\Psi}^t\left(SC_{k}\right)}\setminus S_{n})$)
    \zi \quad\quad\quad\quad\quad $SC_{k}$ unmatch with $S_{n}$;
    %\zi \quad\quad\quad\quad\quad $SC_{k}$-matched-source-node-number-=1;
    %\zi \quad\quad\quad\quad\quad $S_{n}$-matched-SC-number-=1;
    \zi \quad\quad\quad \textbf{else} Refuse the proposal from $S_{m}$;
    \End\End
    \end{codebox}
\item \textbf{Static SD Pair-Subchannel Matching Finished}

    %Jump to power allocation.
\end{enumerate}\\

\hline

\end{tabular}
\end{table}

However, in the matching process, the changes of the source nodes' strategies lead to a dynamic co-channel interference for each source node. As a result, the preference of the source nodes is likely to be changed because of externalities. To adjust the SD pair-subchannel matching with externalities, we will introduce the novel dynamic SD pair-subchannel matching algorithm in the following subsection.
%\footnotetext[1]{A sub-channel is accessible to a source node means that it has never refused the application from the source node before.}
%\footnotetext[2]{The best means the sub-channel that can provide the largest utility, that is, with the best channel gain.}
\subsection{Dynamic SD pair-Subchannel Matching Algorithm}

In this subsection, to develop a sub-channel allocation scheme that fully depicts the interaction between the
SD pairs caused by co-channel interference, we present the dynamic SD pair-subchannel
matching algorithm (DSD-SMA). Different from the SSD-SMA, the preference lists of the source nodes are
adjusted dynamically according to the current matching structure in DSD-SMA. The DSD-SMA contains
a sequence of SSD-SMA iterations, and we do not use a fixed static preference list for each source node
throughout the DSD-SMA process. In each round of the SSD-SMA iteration, it is necessary for each
source node to adjust its static preference list. Because when SSD-SMA is performed, the strategies of
some source nodes will be changed, and the co-channel interference over each sub-channel is also changed
respectively.

The preference lists of the source nodes are adjusted dynamically after a SSD-SMA iteration is performed. In DSD-SMA, the
set of static preference list is extended to $n$ different sets of static preference lists, denoted as
$\textbf{\textit{P}}_s^1, \textbf{\textit{P}}_s^2, \cdots, \textbf{\textit{P}}_s^n$, where $n$ is the index of SSD-SMA iteration in the DSD-SMA. The static preference lists of the source nodes in the $i$th SSD-SMA iteration is given by
$\textbf{\textit{P}}_s^i=\left\{P\left(S_{1}^i\right), P\left(S_{2}^i\right) \cdots, P\left(S_{m}^i\right)\right\}$. For example, $P\left(S_{1}^i\right)=\left\{SC_{P_{i}^{1}}, SC_{P_{i}^{2}}, \cdot, SC_{P_{i}^{K}}\right\}$ means that in the $i$th SSD-SMA iteration, the preference relation satisfies ${SC_{a_{1}} \succ_{S_{1}} SC_{a_{2}} \succ_{S_{1}} \cdots \succ_{S_{1}} SC_{a_{K}}}$. After a SSD-SMA iteration is performed, the matching structure of the network may have been changed and $S_1$'s preference over the set of sub-channels may be different. Player $S_{1}$ then constructs a new static preference list $P\left(S_{1}^{i+1}\right)=\left\{SC_{P_{i+1}^{1}}, SC_{P_{i+1}^{2}}, \cdot, SC_{P_{i+1}^{K}}\right\}$ according to the current matching structure before the $(i+1)$th SSD-SMA iteration. The difference between $\left\{P_{i}^{1}, P_{i}^{2}, \cdots, P_{i}^{K}\right\}$ and $\left\{P_{i+1}^{1}, P_{i+1}^{2}, \cdots, P_{i+1}^{K}\right\}$ is caused by the signal-to-signal interference of the source nodes in the $(i+1)$th SSD-SMA iteration.

To reduce the complexity of the DSD-SMA, we try to avoid the repeated matching proposals in different
SSD-SMA iterations. For a source node $S_i$, there may exist some sub-channels that never accept
source node $S_i$'s proposal in the current matching structure, which have been proved in the past matching
iterations. Hence we define the concept of forbidden pair for the source nodes as follows.

$\textbf{Definition 6:}$ For $S_i\in \cal{S}$ and $SC_j\in \cal{K}$, if
$\bm{\Psi}^t(SC_{j}) \succ _{SC_j} \{\bm{\Psi}^t(SC_{j}) \cup \{S_i\}\}$ and $\phi_{i,j}=0$, $\bm{\Psi}^t(SC_{j})$ is a
\emph{forbidden pair} for $S_i$ over $SC_j$. It is denoted as $\bm{\Psi}^t(SC_{j}) \in F_{S_i}\left(SC_j\right)$.

In this matching model, the strategies of source nodes affect the decision of the sub-channels. For different matching structures, a sub-channel may make different decisions on the same proposal. For example, $SC_k$ will accept $S_i$'s proposal when $\bm{\Psi}^t(SC_{k})=\varnothing$, while $SC_k$ may refuse $S_i$'s proposal when $\bm{\Psi}^t(SC_{k})\neq\varnothing$, if the co-channel interference brought by $S_i$ reduces $SC_k$'s utility. However, the decision of a sub-channel over the same proposal is always the same when the strategies of the source nodes are fixed. The definition of \emph{forbidden pair} shows that the proposal of a source node will not be accepted by a sub-channel if it has been refused with the same strategies of the source nodes before. Each time a proposal of the source node is refused, the current matching structure will be recorded in its \emph{forbidden pair}, and it will not propose itself to the same sub-channel under the same matching structure.

We then describe the whole process of DSD-SMA to solve the sub-channel matching problem with externalities. In each iteration of SSD-SMA, the source nodes will amend their static preference lists with the current matching structure and refresh their \emph{forbidden pairs} each time they are refused by a sub-channel. After one SSD-SMA matching process is performed, it will figure out if the DSD-SMA is over, if not, it will return to the beginning of the SSD-SMA and start the next iteration.

Table~\uppercase\expandafter{\romannumeral4} shows the detailed steps of the DSD-SMA.
The DSD-SMA starts with the process of SSD-SMA, where the source nodes construct their static preference lists of the first
SSD-SMA matching iteration. Then the source nodes that have been matched with less than $q_l$ sub-channels propose themselves to the most preferred sub-channels in their current non-empty static preference lists, and remove the corresponding sub-channels from their current static preference lists. The sub-channels then decide whether to accept the proposals or not according to the criteria of SSD-SMA. If a proposal from source node $S_i$ is refused by a sub-channel $SC_j$, $S_i$ will add the current matching of $SC_j$ into its \emph{forbidden pair}. The process of SSD-SMA completes when no source node is willing to make any proposal with their current static preference lists.

In the next SSD-SMA matching iteration, source nodes first figure out if there exist any \emph{forbidden pairs} with the current matching structure. If there is any, the corresponding sub-channel in the \emph{forbidden pairs} will not be listed in the static preference list of the source node in the following SSD-SMA matching iteration. The set of source nodes then construct their static preference lists of the next SSD-SMA matching iteration according to the current strategies of the source nodes, and the next SSD-SMA matching process is performed. The DSD-SMA process completes when the SSD-SMA matching process performs only one static matching iteration. That is, no source node proposes itself to any sub-channel in the first static matching iteration of the SSD-SMA matching process.

\begin{table}[!ht]
\renewcommand{\arraystretch}{2.0}
\caption{Dynamic SD pair-subchannel matching algorithm (DSD-SMA)}
\centering
\begin{tabular}{p{160mm}}
\hline
\begin{enumerate}
\item \textbf{Matching Iteration}
    \begin{codebox}
    \zi SSD-SMA-matching-process=1;
    \zi \While SSD-SMA-matching-process==1 $\bm{or}$ Static-matching-iteration $>$ 1,
    \zi \quad\For i=1:N
    \zi \quad\quad \For j=1:K
    \zi \quad\quad\quad \If $\Psi^t\left(SC_j\right) \notin F_{S_i}\left(SC_j\right)$
    \zi \quad\quad\quad\quad Calculate $R_{j,i}$;
    \zi \quad\quad r=SSD-SMA-matching-process;
    \zi \quad\quad $S_{i}$ constructs $P_s^{r}\left(S_{i}\right), i\in \cal{S}$;\End\End

    \zi \quad Static-matching-iteration=1;
    \zi \quad Source-node-propose=0;
    \zi \quad\While Static-matching-iteration==1 $\bm{or}$ Source-node-propose=1
    \zi \quad\quad Static-matching-iteration+=1;
    \zi \quad\quad\For i=1:N,
    \zi \quad\quad\If $S_{i}$-matched-SC-number $\leq q_{l}$ $\bm{and}$ $P_s^{r}\left(S_{i}\right) \neq \varnothing$
    \zi \quad\quad\quad $S_{i}$ propose itself to $P_s^{r}\left(S_{i}\right)\left[1\right]$;
    \zi \quad\quad\quad Source-node-propose=1;
    \zi \quad\quad\quad Remove $P_s^{r}\left(S_{i}\right)\left[1\right]$ from $P_s^{r}\left(S_{i}\right)$;\End\End\End\End
    \zi \quad\quad\quad\For i=1:K,
    \zi \quad\quad\quad\quad\If $SC_{k}$ has received any proposal (e.g.$S_{m}$)
    \zi \quad\quad\quad\quad\quad\If $\left\{S_{m}\right\}\cup\bm{\Psi}^t(SC_{k})\succ {}_{SC_{k}}\bm{\Psi}^t(SC_{k})$
    \zi \quad\quad\quad\quad\quad\quad $SC_{k}$ accept the proposal from $S_{m}$;
    %\zi \quad\quad\quad\quad\quad\quad $S_{m}$-matched-SC-number+=1;
    %\zi \quad\quad\quad\quad\quad\quad $SC_{k}$-matched-source-node-number+=1;
    \zi \quad\quad\quad\quad\quad\quad \If $SC_{k}$-matched-source-node-number $\ge q_u$
    \zi \quad\quad\quad\quad\quad\quad\quad \For j=1:N
    \zi \quad\quad\quad\quad\quad\quad\quad\quad \If $S_{j}$ and $SC_{k}$ are matched
    \zi \quad\quad\quad\quad\quad\quad\quad\quad\quad Calculate $F_k({\bm{\Psi}^t\left(SC_{k}\right)}\setminus S_{j})$
    \zi \quad\quad\quad\quad\quad\quad\quad Find the largest $F_k(D)$ in the circulation above (e.g.$F_k({\bm{\Psi}^t\left(SC_{k}\right)}\setminus S_{n})$)
    \zi \quad\quad\quad\quad\quad\quad\quad $SC_{k}$ unmatch with $S_{n}$;
    %\zi \quad\quad\quad\quad\quad\quad\quad $SC_{k}$-matched-source-node-number-=1;
    %\zi \quad\quad\quad\quad\quad\quad\quad $S_{n}$-matched-SC-number-=1;
    \zi \quad\quad\quad\quad\quad\quad\quad Add $\bm{\Psi}^t(SC_{k})$ into $F_{S_n}\left(SC_j\right)$;
    \zi \quad\quad\quad\quad\quad \textbf{else} Refuse the proposal from $S_{m}$;
    \zi \quad\quad\quad\quad\quad\quad Add $\bm{\Psi}^t(SC_{k})$ into $F_{S_m}\left(SC_j\right)$; \End\End\End\End\End
    \end{codebox}
\item \textbf{Matching Finished}

    Jump to power allocation.
\end{enumerate}\\
\hline

\end{tabular}

\end{table}

\subsection{Water Filling Power Allocation}
Power allocation can be implemented after the SD pair-subchannel matching. We divide the power allocation into two phases. In the first phase, the transmitting power of source nodes is allocated through the water filling algorithm, which can be presented as
\begin{eqnarray}
p_{k,m}=\left[\lambda_{k}-\frac{1}{\left|h_{k,m}\right|^{2}{/}\sigma_{s}^{2}}\right]^{+},
\end{eqnarray}
where
\begin{eqnarray}
\lambda_{k}=\frac{1}{\left|W_{m}\right|}\left(P_{SN}+\sum_{i\in {W_{m}}}\frac{1}{\left|h_{k,i}\right|^{2}{/}\sigma_{s}^{2}}\right),
\end{eqnarray}
is the water filling level of $S_{m}$ over $SC_{k}$, and $W_{m}$ is the set of sub-channels allocated to $S_{m}$.

In the second phase, relay $R$ allocates its amplification coefficient over different sub-channels. We assume that the maximum power that relay $R$ allocates to each sub-channel is identical, i.e., $Q_{K}=Q_{R}{/}{K}$. To maximize the sum data rate, relay $R$ provides the maximum power level over every sub-channel, so that the amplification coefficient $G_{k}$ can be given by
\begin{eqnarray}
G_{k}=\sqrt{\frac{Q_{K}}{\sum_{m=1}^{N}p_{k,m}\phi_{k,m}}}.
\end{eqnarray}
\subsection{Stability, Convergence and Complexity }

\subsubsection{Stability and Convergence}

With the definition of \emph{blocking pair} and the \emph{transitive} preference list explained above, we then introduce the
conception of pairwise-stability as below and prove that the proposed SSD-SMA and DSD-SMA both converge to a \emph{pairwise stable} matching.

\textbf{Definition 6:} A matching $\bm{\Psi}^t$ is defined as \emph{pairwise stable} if it is not blocked by any pair which does
not exist in $\bm{\Psi}^t$.

\textbf{Lemma 1:} If the proposed SSD-SMA converges to a matching $\bm\Psi^{*t}$, then
$\bm\Psi^{*t}$ is a \emph{pairwise stable} matching.

\begin{proof}
If $\bm\Psi^{*t}$ is not a \emph{pairwise stable} matching, it means that there exists a pair $\left(S_{m}, SC_{k}\right)_t$,
such that $L\succ_{SC_{k}}\bm{\Psi}^t(SC_{k})$, $L\subseteq\left\{S_{m}\right\}\cup\bm{\Psi}^t(SC_{k})$, $S_{m}\in L$, and
$SC_{k}\succ_{S_{m}}SC_{l}$, $SC_{l}\in\bm{\Psi}^t(S_{m})$. According to the the proposed static SD pair-subchannel matching
algorithm, $S_{m}$ must have proposed itself to $SC_{k}$ before since it can provide a higher utility than $SC_{l}$. We assume
that $SC_{k}$ eliminate $S_{m}$ in the $n$th static matching iteration, denoted as
$\bm{\Psi}^t_{n}(SC_{k})\succ_{SC_{k}}L, L\subseteq \left\{S_{m}\right\}\cup\bm{\Psi}^t_{n}(SC_{k}), S_{m}\in L.$ While $SC_{k}$
only accepts the proposals that provide a larger benefit, we have $\bm{\Psi}^t(SC_{k})\succ_{SC_{k}}\bm{\Psi}^t_{n}(SC_{k})$.
Finally, we have $L\succ_{SC_{k}}\bm{\Psi}^t(SC_{k})$, $\bm{\Psi}^t_{n}(SC_{k})\succ_{SC_{k}}L$, and
$\bm{\Psi}^t(SC_{k})\succ_{SC_{k}}\bm{\Psi}^t_{n}(SC_{k})$, which is contradictory to the \emph{transitive} property of the
preference list. Hence, lemma 1 is proved.
\end{proof}

\textbf{Lemma 2:} If the proposed DSD-SMA converges to a matching $\bm\Psi^{*t}$, then
$\bm\Psi^{*t}$ is a \emph{pairwise stable} matching.

\begin{proof}
If $\bm\Psi^{*t}$ is not a \emph{pairwise stable} matching, it means that there exists a pair $\left(S_{m}, SC_{k}\right)_t$,
such that $L\succ_{SC_{k}}\bm{\Psi}^t(SC_{k})$, $L\subseteq\left\{S_{m}\right\}\cup\bm{\Psi}^t(SC_{k})$, $S_{m}\in L$, and
$SC_{k}\succ_{S_{m}}SC_{l}$, $SC_{l}\in\bm{\Psi}^t(S_{m})$. According to the the proposed DSD-SMA, $S_{m}$ will propose itself
to $SC_{k}$ in the following SSD-SMA matching process since it can
provide a higher utility than $SC_{l}$. The only possibility that $\bm\Psi^{*t}$ is a \emph{pairwise stable} matching is that
$\left\{\bm{\Psi}^t\left(SC_k\right)\right\}$ is a \emph{forbidden pair} of $S_{m}$ over $SC_{k}$, which means that $S_{m}$ has
been refused by $SC_{k}$ under the same matching structure earlier in this DSD-SMA. The
reason $SC_k$ refused $S_{m}$ under this condition is that $\bm{\Psi}^t(SC_{k})$ has a larger product of average throughput than
$\left\{S_{m}\right\}\cup\bm{\Psi}^t(SC_{k})$. It can be denoted as $\bm{\Psi}^t(SC_{k})\succ_{SC_{k}}L$,
$L\subseteq\left\{S_{m}\right\}\cup\bm{\Psi}^t(SC_{k})$, which is contradictory to the assumption. Hence, lemma 2 is proved.
\end{proof}

\textbf{Theorem 2:} The proposed SSD-SMA converges to a \emph{pairwise stable} matching $\bm\Psi^{*}$ after a limited number of static matching iterations in each time slot.

\begin{proof}
As shown in Table \uppercase\expandafter{\romannumeral1}, in each iteration, every source node will propose itself to the
most-preferred sub-channel in its static preference list. No matter the proposal is accepted or not, the source node will remove
this sub-channel from its static preference list and will not propose itself to this sub-channel again. As the
matching goes on, the potential choices for each source node keeps decreasing. So the number of iterations is no more than $K$,
where $K$ is the number of sub-channels, and the proposed SSD-SMA will converge within $K$
iterations. According to Lemma 1, the proposed SSD-SMA converges to a \emph{pairwise stable}
matching.
\end{proof}

\textbf{Theorem 3:} The proposed DSD-SMA converges to a \emph{pairwise stable} matching $\bm\Psi^{*}$ after a limited number of iterations in each time slot.

\begin{proof}
In the process of DSD-SMA, each source node is possible to propose itself to the same sub-channel in different SSD-SMA matching process. However, with the definition of \emph{forbidden pair}, when each time a source node proposes itself to a sub-channel, there are only two possibilities. One possibility is that the sub-channel accepts the proposal and does not disconnect any matched pairs, so that the matched pair of SD pair-subchannel increases. The other possibility is that the sub-channel rejects the matching with a source node, and the disconnected source node will add the current matching into its \emph{forbidden pair}. As the matching goes on, the total number of the matched
pair and \emph{forbidden pair} of each source node over the sub-channels keeps increasing. For each source node, there is no more than $(C_{N-1}^1+C_{N-1}^2+ \cdots +C_{N-1}^{q_u})$ \emph{forbidden pairs} over each sub-channel, where $N$ is the number of source nodes, $q_u$ is the maximum number of SD pairs that can have access to the same sub-channel, so the total number of \emph{forbidden pair} is limited. Because each sub-channel can be matched with no more than $q_{u}$ source nodes, the total number of matched pair is also limited. As a result, the total number of matched pair and \emph{forbidden pair} is limited. The process of DSD-SMA completes with a finite times' proposals. According to Lemma 2, the proposed DSD-SMA converges to a \emph{pairwise stable} matching.
\end{proof}

\subsubsection{Complexity}
In this part we calculate and compare the complexity of the proposed SSD-SMA, DSD-SMA, and the optimal exhaustive search, to
analyse the feasibility of each matching algorithm.

\textbf{Theorem 4:} The complexity of the optimal exhaustive search is $O(2^{KN})$. The iteration number of SSD-SMA is $O(K)$, and the complexity of SSD-SMA is $O(NK^{2})$.

\begin{proof}
For the optimal exhaustive search, relay $R$ exhaustively searches the best subset of SD pairs over every sub-channel. Since every source node and sub-channel can be paired with each other, there exists $K \times N$ possible combinations and the complexity of the optimal exhaustive search is $O(2^{KN})$.

For the SSD-SMA, it contains two phases: the initialization phase and the matching phase.
In the initialization phase, every source node constructs its own static preference list. The initialization of each preference
list is considered as a sorting problem with the complexity of $O(K^2)$, and the total complexity of the initialization phase is
$O(NK^2)$. \footnote{If we also construct the preference list of the sub-channels, it can be proved that there is an extra
complexity of $2^{2N}$. That is why we do not construct the whole preference list of the sub-channels at once.} In the matching
phase, the number of iterations is no more than $K$, and in each iteration, at most $N$ source nodes make proposals, so the
complexity is $O(NK)$. The total complexity of the proposed SSD-SMA is $O(NK^{2})+O(NK)=O(NK^{2})$.
\end{proof}

\textbf{Theorem 5:} The complexity of DSD-SMA in each iteration is $O(NK^{2})$. The upper bound of proposal number in DSD-SMA is $O(KN^{q_u+1})$.

DSD-SMA contains a sequence of SSD-SMA iterations. In each iteration, the SD pairs adjust their preference lists and propose the matching phase of SSD-SMA. The complexity of DSD-SMA in each iteration equals to the complexity of SSD-SMA, that is, $O(NK^{2})$. However, the strict upper bound for the number of iteration in DSD-SMA is hard to obtain, and the reason is given as following. The number of iteration is determined by the relation ship between the adjusted preference list and the matched sub-channels for each SD pair. However, both of the two factors for a SD pair are affected by any other SD pairs. The solution for the strict complexity of DSD-SMA is impeded by the complicated externalities in the network.

To evaluate the complexity of DSD-SMA, we give the upper bound of total proposal number in DSD-SMA. As shown in Theorem 3, each time a SD pair propose itself to a sub-channel, either the total connection number increases or the number of forbidden pair increases. The maximum number of total connection is $K\times q_u$, and the maximum number of total forbidden pair is $K\times N\times C_{N-1}^1+C_{N-1}^2+ \cdots +C_{N-1}^{q_u}$. The upper bound of proposal number in DSD-SMA is $O(K\times q_u)+O(K\times N\times C_{N-1}^1+C_{N-1}^2+ \cdots +C_{N-1}^{q_u})=O(KN^{q_u+1})$.

The whole process of the two-sided resource allocation problem for the NOMA relay network is shown in
Table \uppercase\expandafter{\romannumeral3}.
\begin{table}[!ht]
\renewcommand{\arraystretch}{2.0}
\caption{resource allocation in the NOMA relay network}
\centering
\begin{tabular}{p{160mm}}
\hline
\begin{enumerate}
\item \textbf{SD Pair-Subchannel Matching Algorithm}
    \begin{codebox}
    \zi\quad Perform Static SD Pair-Subchannel Matching Algorithm;
    \zi $\bm{Or}$
    \zi\quad Perform Dynamic SD Pair-Subchannel Matching Algorithm;
    \end{codebox}
\item \textbf{Power Allocation}
    \begin{codebox}
    \zi \For m=1:N
    \zi \quad \For k=1:K
    \zi \quad\quad $S_i$ allocates its transmitting power over $SC_k$;
    \zi \quad\quad $\lambda_{k}=\frac{1}{\left|W_{m}\right|}\left(P_{SN}+\sum_{i\in {W_{m}}}\frac{1}{\left|h_{k,i}\right|^{2}{/}\sigma_{s}^{2}}\right)$;
    \zi \quad\quad $p_{k,m}=\left[\lambda_{k}-\frac{1}{\left|h_{k,m}\right|^{2}{/}\sigma_{s}^{2}}\right]^{+}$;\End\End
    \zi i=1:K
    \zi \quad Relay $R$ allocates the amplification factor of $SC_k$;
    \zi \quad $G_{k}=\sqrt{\frac{Q_{K}}{\sum_{m=1}^{N}p_{k,m}\phi_{k,m}}}$;
    \end{codebox}
\item \textbf{Resource Allocation Finished}

\end{enumerate}\\
\hline

\end{tabular}
\end{table}

\section{Simulation Results}

In this section, we evaluate the performance of the proposed SSD-SMA and DSD-SMA in NOMA scheme considering proportional fair.
We compare the proposed algorithms with the OFDMA scheme and the optimal exhaustive search. In the OFDMA scheme, each sub-channel can only be allocated to one SD pair at one time, while one SD pair may have access to multiple sub-channels. For the optimal exhaustive search, each source node and each sub-channel can be paired with any number of players from the opposite side as long as they want.

In the simulation, most of the parameters are set based on the existing LTE/LTE-Advanced specifications~\cite{3GPPTR} \cite{3GPPTS}. The radio resource allocation is updated every 1 ms, and the user throughput averaged over 10 ms is measured. We set $q_u$ and $q_l$ as 8 and 3 respectively and all curves are generated by averaging over 1000 instances of the algorithms.
Table \uppercase\expandafter{\romannumeral4} is the parameters of the simulation.
%In the simulation, we set the peak power of each source node $P_{SN}$ as 46 dBm, and the maximum power of relay $R$, $Q_R$ as
%86 dBm so that the maximum power over each sub-channel $Q_K$ is approximately around 46 dBm. The noise variance $\sigma^2$ is -174
%dBm , the path loss coefficient $\alpha$ is set as 3.76, and all user nodes are uniformly distributed in an square area with the
%size of length 200 m, with relay $R$ in the center of the square. Since the sub-channels are equally distributed from the spectrum
%resource, we omit the parameter of bandwidth in the simulation and the results are shown in every Hz. The radio resource allocation
%is updated every 1 ms, and the user throughput averaged over 10 ms is measured. We obtain the simulation results as shown below,
%and all curves are generated based on averaging over 1000 instances of the algorithms. We set $q_u$ and $q_l$ as 8 and 3
%respectively considering the number of sub-channels and SD pairs. Table \uppercase\expandafter{\romannumeral6} is the parameters
%of the simulation.

\begin{table}[!ht]
\centering
\caption{Simulation Parameters}
\begin{tabular}{|c|c|}
\hline
\textbf{Parameter} & \textbf{Value}\\
\hline
cell range & 200m square\\
\hline
number of sub-channels (K) & 10\\
\hline
number of SD pairs (N) & 5 to 50\\
\hline
peak power of source node & 46 dBm\\
\hline
peak power of relay & 86 dBm\\
\hline
noise & additive white Gaussian noise\\
\hline
noise variance ($\sigma^2$) & -174 dBm\\
\hline
path loss coefficient $\alpha$ & 3.76\\
\hline
center frequency & 2GHz\\
\hline
bandwidth & 4.5MHz\\
\hline
scheduling interval & 1 ms\\
\hline
averaging interval of user throughput & 10 ms\\
\hline
$q_{u}$ & 8\\
\hline
$q_{l}$ & 3\\
\hline
scheduler & proportional fairness\\
\hline
fading & Rayleigh fading\\
\hline
\end{tabular}
\end{table}

\begin{figure*}
\begin{minipage}{0.45\linewidth}
\centerline{\includegraphics[width=3.2in]{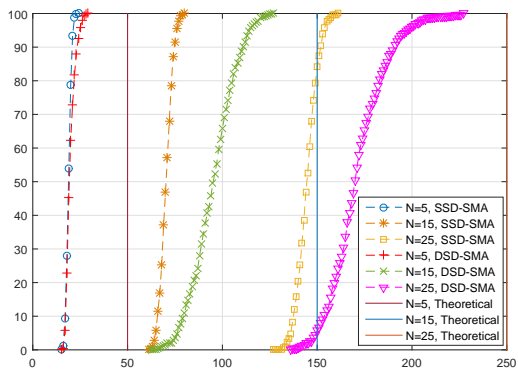}}
\centerline{(a)}
\end{minipage}
\hfill
\begin{minipage}{0.45\linewidth}
\centerline{\includegraphics[width=3.2in]{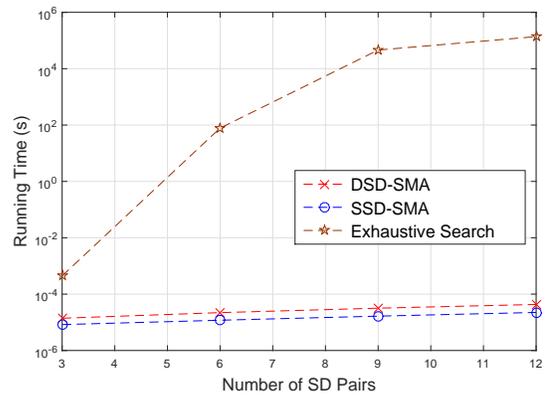}}
\centerline{(b)}
\end{minipage}
\caption{Number of SD-pairs vs. Number of Static Matching Iteration \& Running Time.}
\vspace{-8mm}
\label{dlp}
\end{figure*}

Fig.2 (a) shows the number of proposals vs. CDF of the number of proposals with 5, 15 and 25 SD pairs in different matching algorithms. It is illustrated that the number of proposal increases with the increment of the SD pair number. The CDF curves for SSD-SMA is steeper than DSD-SMA, which means that the number of static iterations in SSD-SMA is more stable than DSD-SMA. It is also shown that the number of iteration for SSD-SMA is smaller than DSD-SMA with the same number of SD pairs. The difference between two algorithms becomes larger with more SD pairs, which is determined by the nature of the algorithms. We also give the theoretical curve for the largest proposal number of SSD-SMA in the matching phase, which agree with our simulation curve.

In Fig.2 (b), we simulate the number of SD pairs vs. the running time with different matching algorithms. The running time of each matching algorithm reflects its complexity and is shown in exponential form. The running time of SSD-SMA and DSD-SMA grow gradually with the increment of SD pair, and the running time of DSD-SMA is always a little bit larger than that of the SSD-SMA. The running time of the exhaustive search increase rapidly with the number of SD pairs, and is much larger than the proposed algorithms, which further proved its infeasibility.

\begin{figure}[!ht]
\centerline{\includegraphics[width=10cm]{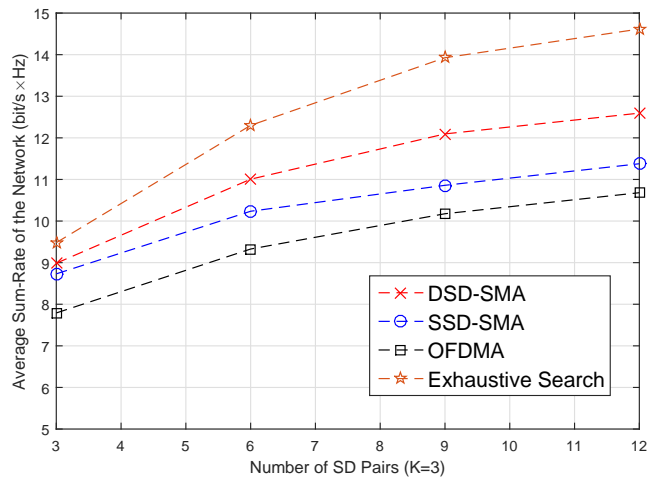}}
\caption{Number of SD-pairs vs. Average sum-rate of the network.}
\end{figure}

%\begin{figure}[!ht]
%\centerline{\includegraphics[width=12cm]{fig1.eps}}
%\vspace{-1mm}
%\caption{Number of SD-pairs vs. Total average sum-rate.}
%\vspace{-3mm}
%\end{figure}

Fig.3 illustrates the relation between average sum-rate and the number of SD pairs and shows the comparison of average sum-rate of the proposed NOMA schemes, OFDMA scheme, and the optimal exhaustive search. As proved in Section \uppercase\expandafter{\romannumeral3}.E, the complexity of the optimal exhaustive search increases exponentially with the number of SD pairs and sub-channels. To get the simulation result of the optimal exhaustive search in a regular time, we have to decrease the number of SD pairs and sub-channels in the simulation. Here we set the number of sub-channels as 3, and the number of SD pairs is reduced to 3-12, while other parameters are still the same as in Table \uppercase\expandafter{\romannumeral4}. The curve of optimal exhaustive search is generated based on averaging over 100 instances of the algorithms.

The performance of the proposed DSD-SMA and SSD-SMA in NOMA scheme outperform the OFDMA scheme significantly because it achieves a more efficient utilization of spectrum resource. The average sum-rate increases as the number of SD pair grows, and the growth becomes slower as $N$ turns larger because of the saturation of channel capacity. But when the number of SD pairs is much larger than the number of sub-channels, the average sum-rate continues to increase at a low speed due to the multiuser gain. When comparing the SSD-SMA with DSD-SMA, we can observe that with the increment of SD pairs, the advantage of DSD-SMA over SSD-SMA in the average sum-rate of the network becomes larger. That is because when the network gets more crowded, the impact of the externalities will be more significant. DSD-SMA can adjust its matching based on the externalities and provide a larger average sum-rate.

It can be noted the DSD-SMA can reach almost $95\%$ of the average sum-rate of the optimal exhaustive search when there are only 3 SD pairs in the network. The average sum-rate of the SSD-SMA is close to that of the DSD-SMA and the optimal exhaustive search with 3 SD pairs in the network. However, the performance gap of the proposed algorithms to the optimal exhaustive search increases with the increment of SD pairs. It makes sense since the complexity of the optimal exhaustive search is significantly more than that of the proposed algorithms, as the number of SD pair increases. It can also be noted that the proposed algorithm can always achieves a significantly higher sum rate compared to the OFDMA scheme.
\vspace{-6mm}
\begin{figure}[!ht]
\centerline{\includegraphics[width=10cm]{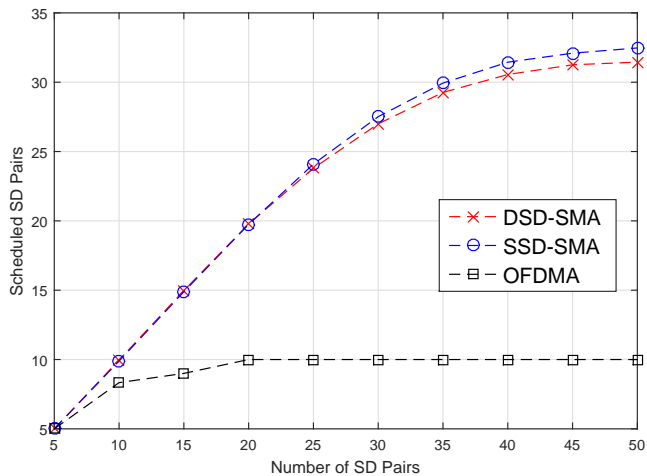}}
\vspace{-5mm}
\caption{Number of SD-pairs vs. Number of Scheduled Users.}
\vspace{-6mm}
\end{figure}

Fig.4 depicts the number of scheduled SD pairs vs. the number of SD pairs in different matching structures. The scheduled SD pairs in OFDMA scheme is no more than the number of sub-channels since each sub-channel can be matched with no more than one SD pair. In NOMA scheme, the number of
scheduled SD pair increases almost linearly when there are less than $\frac{K \times q_u}{q_l}$ SD pairs, because there are plenty of spare spectrum resources. When $N$ increases, the number of scheduled SD pair becomes saturated and increases slowly due to the multiuser gain. SSD-SMA slightly outperforms DSD-SMA in terms of the number of scheduled SD pairs when the network gets crowded.

%\begin{figure}[!ht]
%\centerline{\includegraphics[width=12cm]{fig2.eps}}
%\vspace{-1mm}
%\caption{Number of SD-pairs vs. Average rate of each scheduled SD pair.}
%\vspace{-3mm}
%\end{figure}

%Fig.6 depicts the number of SD pairs vs. the average rate of each scheduled SD pair in different matching conditions. By comparing the curves, we find that the average rate of each scheduled SD pair decrease rapidly when there are few SD pairs in this network because of the limited spectrum and power resource in this network. But the tendency of the average rate turns different in different access schemes when the network is crowded. The average rate keeps decreasing with a slower trend in NOMA scheme while it turns to increasing slowly in OFDMA scheme due to different matching principles. In NOMA scheme, more SD pairs will share the resource of the network with the increment of SD pair number. In OFDMA scheme, each sub-channel can only be matched with one SD pair. When the network is crowded, the number of scheduled SD pair barely change with the increment of total SD pairs in the network, and the average rate of each scheduled SD pair will increase slowly due to the diversity channel gain. The tendency of dynamic SD pair-subchannel matching algorithm and static SD pair-subchannel matching algorithm are similar, but dynamic SD pair-subchannel matching algorithm outperforms static SD pair-subchannel matching algorithm weakly in average rate.
%\vspace{-6mm}
\begin{figure}[!ht]
\centerline{\includegraphics[width=10cm]{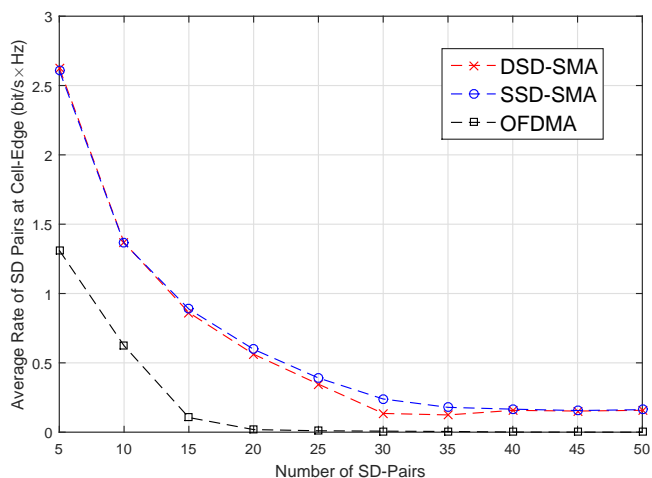}}
\vspace{-5mm}
\caption{Number of SD-pairs vs. Average rate of each cell-edge SD pair.}
\vspace{-8mm}
\end{figure}

Fig.5 shows the number of SD pairs vs. the average rate of each cell-edge SD pair in the network in different matching structures. We define the total distance of a SD pair as the distance between the source node and relay $R$ plus the distance between relay $R$ and the destination node. The cell-edge SD pairs are those with a total distance of more than 160 m. In the matching with the proposed algorithms, the average rate of cell-edge SD pairs outperform the OFDMA scheme obviously. That is because in each time slot, there are more access SD pairs in NOMA scheme than in OFDMA scheme, and the cell edge users have a larger chance to access the channels in NOMA scheme. The cell edge users have a relatively high average rate when there are only a few SD pairs in the network. The average rate decreases rapidly as the number of SD pairs becomes larger because of the limited spectrum resource. The average rate of the cell-edge SD pairs turn to be stable in a crowded network. The only difference between SSD-SMA and DSD-SMA lies in the middle of the curve, where the average rate of the cell edge SD pair in DSD-SMA is slightly lower than that in the SSD-SMA. In OFDMA scheme, the average rate of cell-edge SD pair tend to be very low when there are over 10 SD pairs in the network due to its small number of accessed SD pairs.
\vspace{-6mm}
\begin{figure}[!ht]
\centerline{\includegraphics[width=10cm]{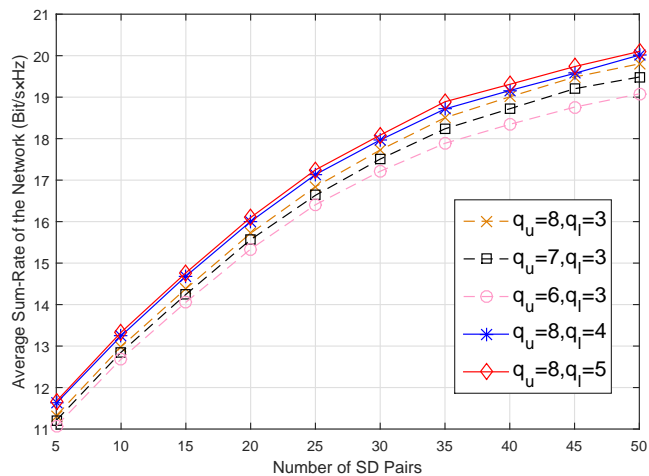}}
\vspace{-5mm}
\caption{Number of SD-pairs vs. Average sum-rate with different $q_{u}$ and $q_l$ in DSD-SMA.}
\vspace{-8mm}
\end{figure}

Fig. 6 shows the average sum-rate vs. the number of SD pairs for different $q_u$ and $q_l$ in the proposed DSD-SMA. When $q_l$ is fixed, it can be seen that the average sum-rate is higher with a larger $q_u$. With $q_u$ fixed, the average sum-rate also increases with $q_l$. When $q_u$ increases, the marginal increment of the average sum-rate becomes smaller. Due to the increased inter-signal interference, it is more difficult for a new coming SD pair to further enhance the average sum-rate over the sub-channels. As $q_l$ increases, the marginal increment of the average sum-rate also becomes smaller as the possibility that a matching is denied by the restriction of $q_l$ decreases.
\vspace{-5mm}
\section{Conclusion}
\vspace{-2mm}
In this paper, we studied the resource allocation problem in a NOMA wireless network with a one-way OFDM AF relay by optimizing
the sub-channel assignment and the power allocation. By formulating the problem as a many-to-many two-sided matching problem, we
proposed two near optimal SD pair-subchannel matching algorithms in which the SD pairs and sub-channels can be matched and converge to a stable matching. The average sum-rate of the proposed SSD-SMA and DSD-SMA in NOMA scheme are higher than that of the conventional OFDMA network and close to that of optimal exhaustive search. The proposed matching algorithms can serve more users and provide better service to the cell-edge users when compared with the conventional OFDMA scheme. The proposed SSD-SMA has a lower complexity while DSD-SMA performs better in average sum-rate of the whole network.
\vspace{-3mm}
\begin{appendices}
\vspace{-2mm}
\section{}
Proof of \textbf{Theorem 1}. The sum-rate maximization problem in (11a) is NP-hard.
\begin{proof}
The proof of this theorem can be divided into two parts, $q_u=1$ and $q_u>1$.

When $q_u=1$, $\left( \ref{system_1} \right)$ becomes an joint power and sub-channel allocation problem in the traditional OFDMA
system, which has been proved to be NP-hard in \cite{LD2014}.

When $q_u>1$, we proof that the problem is NP-hard even when we omit the power allocation problem and allocate the power equally
over each sub-channel. Since the 3-dimensional matching problem (3-DM problem) has been proven to be NP-complete in \cite{K1991}, we try to construct a instance which can be proved to be equal with a 3-DM problem. When the decision problem of this specific instance is proved to be NP-complete, the instance of $\left( \ref{system_1} \right)$ with equal power allocation is an NP-hard problem \cite{S2012}.

We construct a instance where $q_u=2$ and $q_l=1$. Suppose that relay R allocates the power equally over each sub-channel for the accessed SD pairs, and the SD pairs are separated into two disjoint sets $\cal{M}_\text{1}$ and $\cal{M}_\text{2}$ such that
$\cal{M}_\text{1} \cap \cal{M}_\text{2}=\varnothing$ ,$\cal{M}_\text{1} \cup \cal{M}_\text{2}$ is the whole set of SD pairs
$\cal{S} \cup \cal{D}$, and $\left|\cal{M}_\text{1}\right|=N/2, \left|\cal{M}_\text{2}\right|=N/2,$ and $\left|\cal{K}\right|=K$.
For each sub-channel, we assume that it is allocated to one SD pair from $\cal{M}_\text{1}$ and the other from $\cal{M}_\text{2}$.
Let $\cal{Q}$ be a collection of ordered triples $\cal{Q} \subseteq \cal{K}\times\cal{M}_\text{1}\times\cal{M}_\text{2}$, where
$\mathbf{Q}_\text{i}$ = $\left(SC_{i}, M_{i}, M_{j}\right)\in \cal{Q}.$ According to $\left( \ref{system_0} \right)$, the sum-rate
of any triple $\mathbf{Q}_\text{i}$ can be set as $R_{Q_i}$. We need to determine whether there exists a set
$\cal{Q}' \subseteq \cal{Q}$ so that $\left|\cal{Q}'\right|$ = $\min \left\{N/2, K\right\}$,
$\sum_{i=1}^{\min \left\{N/2, K\right\}}{U_{\cal{Q}'_\text{i}}}\leq \lambda$, where any $\mathbf{Q}'_\text{i} \in \cal{Q}'$ and
$\mathbf{Q}'_\text{i} \in \cal{Q}'$ do not contain the same components. For $\cal{Q}' \subseteq \cal{Q}$, it is a 3-DM if the
followings hold: $\left(1\right)$ $\left|\cal{Q}'\right|$ = $\min \left\{N/2, K\right\}$; $\left(2\right)$ For any two distinct
triples, $\left(SC_{i}, M_{i}, M_{j}\right)\in \cal{Q}$ and $\left(SC_{p}, M_{p}, M_{q}\right)\in \cal{Q}'$, we have
$i \neq j \neq p \neq q$. If we set $\lambda$ to an infinite negative, the problem we formed will be reduced to a 3-DM decision
problem. Therefore, the decision problem of this instance is NP-complete, and the corresponding instance is NP-hard.
Since a special case of $\left( \ref{system_1} \right)$ is proved to be NP-hard, the sum-rate maximization problem in~$\left( \ref{system_1} \right)$ is NP-hard.
\end{proof}
\end{appendices}


\vspace{-5mm}
\begin{thebibliography}{4}
\bibitem{ZDSL2016}
S. Zhang, B. Di, L. Song, and Y. Li, ``Radio Resource Allocation for Non-orthogonal Multiple Access (NOMA) Relay Network Using
Matching Game," in\emph{ Proc. IEEE Int. Commun. Conf.,} pp. 1-6, Kuala Lumpur, May 2016.
\bibitem{DWYHIW2015}
L. Dai, B. Wang, Y. Yuan, S. Han, C. I, and Z. Wang, ``Non-orthogonal Multiple Access for 5G: Solutions, Challenges, Opportunities, and Future Research Trends," \emph{IEEE Commun. Mag.,} vol. 53, no. 9, pp. 74-81, Sep. 2015.
\bibitem{TGWIJFA2014}
J. Thompson, X. Ge, H. Wu, R. Irmer, H. Jiang, G. Fettweis, and S. Alamouti, ``5G Wireless Commnucation Systems:
Prospects and Challenges," \emph{IEEE Commun. Mag.,} vol. 52, no. 2, pp. 62-64, Feb. 2014.
\bibitem{YYYHLLT2016}
Y. Yuan, Z. Yuan, G. Yu, C-H. Hwang, P-K. Liao, A. Li, and K. Takeda, ``Non-orthogonal transmission technology in LTE evolution," \emph{IEEE Commun. Mag.,} vol. 54, no. 7, pp. 68-74, Jul. 2016.
\bibitem{HB2015}
K. Higuchi and A. Benjebbour, ``Non-orthogonal Multiple Access(NOMA) with Successive Interference Cancellation for Future Radio
Access," \emph{IEICE Trans. Commun.,} vol. E98-B, no. 3, Mar. 2015.
%\bibitem{PLWL2006}
%L. Ping, L. Liu, K. Wu, and W. Leung, \emph{``Interference Division Multiple-Access,¡±} in IEEE Trans. Wireless Commun., vol. 5, no. 4, pp. 938-947,
%Apr. 2006.
%\bibitem{IIT2012}
%M. Imari, M. Imran, and R. Tafazolli, \emph{``Low Density Spreading Multiple Access,¡±} in J. Inf. Technol. Softw. Eng., vol. 2, no. 4, pp. 1-2, Sep. %2012.
\bibitem{LMZ2014}
J. Liberti, S. Moshavi, and P. Zablocky, ``Successive interference cancellation," U.S. Patent 8670418 B2, Mar. 2014.
%\bibitem{LHK2014}
%A. Li, A. Harada, H. Kayama, \emph{``A Novel Low Computational Complexity Power Assignment Method for Non-orthogonal Multiple Access Systems,¡±} in IEICE Trans. Fundam. Electron. Commun. Comput. Sci., pp. 57-68, vol. E97-A, no .1, Jan. 2014.
\bibitem{YHBLLJ2015}
C. Yan, A. Harada, A. Benjebbour, Y. Lan, A. Li, and H. Jiang, ``Receiver Design for Downlink Non-Orthogonal Multiple Access (NOMA),"
in\emph{ Proc. IEEE Veh. Tech. Conf.,} pp. 1-6, Glasgow, May 2015.
%\bibitem{SBKN2013}
%Y. Saito, A. Benjebbour, Y. Kishiyama, and T. Nakamura, ``System-Level Performance Evaluation of Downlink Non-orthogonal Multiple Access (NOMA)," in \emph{Proc. IEEE Int. Symp. on Personal Indoor and Mobile Radio Commun. Conf.,} pp. 611-615, London, UK, Sep. 2013.
%\bibitem{TK2015}
%S. Timotheou, and I. Krikidis, ``Fairness for Non-Orthogonal Multiple Access in 5G Systems," \emph{IEEE Signal Process. Lett.,}
%vol. 22, no. 10, pp. 1647-1651, Oct. 2015.
%\bibitem{IXIT2014}
%M. Al-Imari, P. Xiao, M. A. Imran, and R. Tafazolli, ``Uplink Non-Orthogonal Multiple Access for 5G Wireless Networks," in \emph{ Proc. IEEE Int. Symp. on Wireless Commun. Syst. Conf.,} pp. 781-785, Barcelona, Spain, Aug. 2014.
%\bibitem{OKH'2012}
%N. Otao, Y. Kishiyama, and K. Higuchi, ``Performance of Non-orthogonal Access with SIC in Cellular Downlink Using Proportional Fair-Based Resource Allocation," in\emph{ Proc. IEEE Int. Symp. on Wireless Commun. Syst. Conf.,} pp. 476-480, Paris, France, 2012.
%\bibitem{CBLH2014}
%X. Chen, A. Benjebbour, A. Li, and A. Harada, ``Multi-User Proportional Fair Scheduling for Uplink Non-orthogonal Multiple Access (NOMA)," in\emph{ Proc. IEEE Veh. Technol. Conf.,} pp. 1-5, Seoul, Korea, May 2014.

%\bibitem{C2014}
%J. Choi, ``Non-Orthogonal Multiple Access in Downlink Coordinated Two-Point Systems," \emph{IEEE Signal Process. Lett.,} vol. 18, no. 2, pp. 313-316, Feb. 2014.
%\bibitem{KL2015'}
%J. Kim and I. Lee, ``Non-Orthogonal Multiple Access in Coordinated Direct and Relay Transmission," \emph{IEEE Commun. Lett.,} vol. 19, no. 11, pp. 2037-2040, Nov. 2015.
%\bibitem{LBLH2014}
%Y. Lan, A. Benjebbour, A. Li, and A. Harada, ``Efficient and Dynamic Fractional Frequency Reuse for Downlink Non-orthogonal
%Multiple Access," in\emph{ Proc. IEEE Veh. Technol. Conf.,} pp. 1-5, Seoul, Korea, May 2014.
%\bibitem{HFND2015}
%M. Hojeij, J. Farah, C. A. Nour, and C. Douillard, ``Resource Allocation in Downlink Non-orthogonal Multiple Access (NOMA) for Future Radio Access" in\emph{ Proc. IEEE Veh. Technol. Conf.,} pp. 1-6, Glasgow, Britain, May 2015.
%\bibitem{DFP2015}
%Z. Ding, P. Fan, and H. V. Poor, ``Impact of User Pairing on 5G Non-Orthogonal Multiple Access Downlink Transmissions", \emph{IEEE
%Trans. Veh. Technol.,} vol. pp, no. 99, pp. 1-14, Sep. 2015.
\bibitem{DBSL2015}
B. Di, S. Bayat, L. Song, and Y. Li, ``Radio Resource Allocation for Downlink Non-Orthogonal Multiple Access (NOMA)
Networks using Matching Theory," in \emph{Proc. IEEE Global Commun. Conf.,} pp. 1-6, San Diego, Dec. 2015.

\bibitem{MG2015}
J. Men and J. Ge, ``Non-Orthogonal Multiple Access for Multiple-antenna Relaying Networks,"\emph{ IEEE Commun. Lett.,} vol. 19, no. 10, pp. 1686-1689, Oct. 2015.
\bibitem{MVB2015}
A. Mohamad, R. Visoz, and A. O. Berthet, ``Code Design for Multiple-Access Multiple-Relay Wireless Channels with Non-Orthogonal
Transmission," in \emph{Proc. IEEE Int. Conf. Commun.,} pp. 2318-2324, London, Jun. 2015.
\bibitem{LDEP2016}
Y. Liu, Z. Ding, M. Elkashlan, and H. V. Poor, ``Cooperative Non-Orthogonal Multiple Access with Simultaneous Wireless Information and Power Transfer," \emph{IEEE J. Sel. Areas Commun.,} vol. 34, no. 4, pp. 938-953, Mar. 2016.
\bibitem{ZMXDF2016}
Z. Zhang, Z. Ma, M. Xiao, Z. Ding, and P. Fan, ``Full Duplex Device-to-Device Aided Cooperative Non-Orthogonal Multiple Access," \emph{IEEE Trans. Veh. Tech.} vol. pp, no. 99. pp. 1-1. Aug. 2016.
\bibitem{DDP2016}
Z. Ding, H. Dai, and H. V. Poor, ``Relay Selection for Cooperative NOMA," \emph{IEEE Wireless Commun. Lett.,} vol. 5, no. 4, pp. 416-419, Jun. 2016.

\bibitem{RS1992}
A. Roth and M. Sotomayor, \emph{``Two-Sided Matching: A Study in Game Theoretic Modeling and Analysis,"} Cambridge, UK: Cambridge Univ. Press, 1992.
\bibitem{PY2015}
M. Pycia and M. B. Yenmez, ``Matching with Externalities," \emph{Socail Science Electronic Publishing,} May, 2015.
\bibitem{GHLP2009}
R. H. Gohary, Y. Huang, Z. Q. Luo, and J. S. Pang, ``A Generalized Iterative Water-Filling Algorithm for Distributed Power Control in the Presence of a Jammer,"\emph{ IEEE Trans. Signal Process.,} vol. 57, no. 7, pp. 2660-2674, Feb. 2009.
\bibitem{SKBNLH2013}
Y. Saito, Y. Kishiyama, A. Benjebbour, T. Nakamura, A. Li, and K. Higuchi, ``Non-orthogonal Multiple Access (NOMA) for Cellular
Future Radio Access," in\emph{ Proc. IEEE Veh. Tech. Conf.,} pp. 1-5, Dresden, Jun. 2013.
\bibitem{BLHVL-2014}
S.~Bayat, R.~Louie, Z.~Han, B.~Vucetic and Y.~Li, ``Distributed User Association and Femtocell Allocation in Heterogeneous Wireless Networks," \emph{IEEE Trans. Commun.}, pp.~3027-3043, vol.~62, no.~8, Aug.~2014.
\bibitem{NAYLM2016}
H. Q. Ngo, A. Ashikhmin, H. Yang, E. G. Larsson, and T. L. Marzetta, ``Cell-free massive MIMO versus small cells," \emph{IEEE Trans. Wireless Commun.,} 2016, submitted.
\bibitem{LJJZM2016}
P. Liu, S. Jin, T. Jiang, Q. Zhang, and M. Matthaiou, ``Pilot power allocation through user grouping in multi-cell massive MIMO systems," \emph{IEEE Trans. Commun.,} 2016, submitted.
\bibitem{UKH-2012}
J.~Umehara, Y.~Kishiyama, and K.~Higuchi, ``Enhancing User Fairness in Non-orthogonal Access with Successive Interference
Cancellation for Cellular Downlink," in\emph{ Proc. Inter. Conf. Commun. System,} pp. 1-5, Omaha, Nov.~2012.
\bibitem{T2005}
D. Tse, ``Downlink AWGN Channel" in \emph{Fundamentals of Wireless Communication,} Cambridge, UK: Cambridge Univ. Press, 2005, ch.~6, sec. 2, pp. 238-242.
\bibitem{WCYW2014}
S. Wang, L. Chen, Y. Yang, and G. Wei, ``Quantization in Uplink Multi-Cell Processing with Fixed-order Successive Interference
Cancellation Scheme under Backhaul Constraint," in\emph{ Proc. IEEE Veh. Tech. Conf.,} pp. 1-5, Seoul, May~2014.
\bibitem{OKH2012}
N. Otao, Y. Kishiyama, and K. Higuchi, ``Performance of non-orthogonal access with SIC in cellular downlink using proportional
fair-based resource allocation," in\emph{ Proc. IEEE Int. Symp. Wireless Commun. System,} pp. 476-480, Paris, Aug. 2012.
%\bibitem{CALA2014}
%X. Chen, A. Benjebbour, A. Li, and A. Harada, ``Multi-User Proportional Fair Scheduling for Uplink Non-Orthogonal Multiple Access (NOMA)," in\emph{ Proc. IEEE Veh. Technol. Conf.,} pp. 1-5, Seoul, Korea, May 2014.
\bibitem{KG2005}
M. Kountouris and D. Gesbert, ``Memory-based opportunistic multi-user beamforming," in\emph{ Proc. IEEE Int. Symp. Inform. Theory,} pp. 1426-1430, Adelaide, Sept. 2005.
\bibitem{WN2014}
L. Wolsey and G. Nemhauser, \emph{``Integer and Combinatorial Optimization,"} John Wiley $\&$ Sons, USA, 2014.
\bibitem{BLVL2013}
S. Bayat, R. Louie, B. Vucetic, and Y. Li, ``Dynamic decentralised algorithms for cognitive radio relay networks with multiple primary and secondary users utilising matching theory," \emph{Trans. Emerging Telecommun. Technol.,} vol. 24, no. 5, pp. 486-502, May, 2013.
\bibitem{BLLV2011}
S. Bayat, R. Louie, Y. Li and B. Vucetic, ``Cognitive radio relay networks with multiple primary and secondary users: distributed stable matching algorithms for spectrum access," in \emph{Proc. IEEE Int. Conf. Commun.,} pp. 1-6, Kyoto, Jun. 2011.
\bibitem{LJZH2011}
P. Lin, J. Jia, Q. Zhang, and M. Hamdi, ``Dynamic Spectrum Sharing With Multiple Primary and Secondary Users," \emph{IEEE Trans. Veh. Tech.,} vol. 60, no. 4, pp. 1756-1765, Mar. 2011.
\bibitem{HH2015}
M. Hasan, and E. Hossain, ``Distributed Resource Allocation for Relay-Aided Device-to-Device Communication Under Channel Uncertainties: A Stable Matching Approach," \emph{IEEE Trans. Commun.,} vol. 63, no. 10, pp. 3882-3897, Aug.~2015.
\bibitem{YRBC2004}
W. Yu, W. Yhee, S. Boyd, and J. Cioffi, ``Iterative Water-Filling for Gaussian Vector Multiple-Access Channel," \emph{IEEE Trans. Inf. Theory,} pp. 322, vol. 50, no. 1, Jan. 2004.
\bibitem{3GPPTR}
3GPP TR 25.996, ``Spatial channel model for Multiple Input Multiple Output (MIMO) simulations," Release 12, Sept.~2014.
\bibitem{3GPPTS}
3GPP TS 36.213, ``Evolved Universal Terrestrial Radio Access (E-UTRA) Physical Layer Procedures," Release 12, Sept. 2014.
\bibitem{LD2014}
Y. Liu and Y. Dai, ``On the Complexity of Joint Subcarrier and Power Allocation for Multi-user OFDMA Systems," \emph{IEEE Tran. Signal Process.,} vol. 62, no. 3, pp. 583-596, Feb. 2014.
\bibitem{K1991}
V. Kann, ``Maximum Bounded 3-dimensional Matching is MAX SNP-complete," \emph{IEEE Info. Process. Lett.,} vol. 37, no. 1, pp. 27-35, Jan. 1991.
\bibitem{S2012}
M. Sipser, ``Introduction to the Theory of Computation," Cengage Learning, USA, 2012.

%\bibitem{HSD2014}
%K. Hamidouche, W. Saad, and M. Debbah, ``Many-to-Many Matching Games for Proactive Social-Caching in Wireless Small Cell Networks," in\emph{ Proc. Int. Symp. Modeling and Optimization in Mobile, Ad Hoc, and Wireless Networks (WiOpt),} pp. 569-574, Hammamet, Tunisia, May 2014.

\end{thebibliography}
\end{document}